%%%%%%%%%%%%%%%%%%%%%%%%%%%%%%%%%%%%%%%%%%%%%%%%%%%%%%%%%%%%%%%%%%%%%%%%%%%%%%%
%% Page layout:
\magnification = \magstephalf
\baselineskip = 13truept
\vsize = 9.1truein
%\voffset = -0.05truein
\nopagenumbers
\raggedbottom
\overfullrule=0pt
\def\status{To Appear in {\it The Astrophysical Journal}}
\headline = {\tenrm\ifnum \pageno > 1 \centerline{-- \folio\ --}\else\centerline{\status}\fi}
%\footline = {\tenrm\ifnum \pageno = 1 \centerline{\status}\else\hfil\fi}
\def\sec#1 #2{\bigskip\medskip\centerline{\apj {#1}~{#2}}}
\def\subsec#1 #2{\bigskip\centerline{{#1}~{\sl #2}}\medskip}
\def\subsubsec#1 #2{\medskip\centerline{{#1}~{#2}}\medskip}
\def\fpsec{\bigskip}

%% Fonts:

 at 12truept

\font\apj = cmcsc10

%% Definitions:
\def\et{et~al.\ }
\def\eg{e.g.,\ }
\def\ie{i.e.,\ }
\def\hi {\noindent \hangindent=2.5em}
\def\nts{\negthinspace}

\def\LX{{\sl L}$_{\rm X}$}
\def\FX{{\sl F}$_{\rm X}$}
\def\LF{{\sl L}$_{\rm FIR}$}

\def\E#1{$10^{#1}\, {\rm ergs\> s^{-1}}$}
\def\EE#1 #2{$#1 \times 10^{#2}\, {\rm ergs\> s^{-1}}$}
\def\FF#1 #2{$#1 \times 10^{#2}\, {\rm ergs\> cm^{-2}\> s^{-1}}$}

\def\Ha{H$\alpha $}
\def\Hb{H$\beta $}
\def\HII {H~{\apj ii}}
\def\OI {[O~{\apj i}]}
\def\OII {[O~{\apj ii}]}
\def\OIII {[O~{\apj iii}]}
\def\NI {[N~{\apj i}]}
\def\NII {[N~{\apj ii}]}
\def\SII {[S~{\apj ii}]}

\def\Msun{\ifmmode M_{\odot} \else $M_{\odot}$\fi}

\def\errora#1 #2 #3{$#1^{\scriptscriptstyle
+#2}\!\!\!\!\!\!\!\!\!_{\scriptscriptstyle -#3}$}
\def\errorb#1 #2 #3{$#1^{\scriptscriptstyle
+#2}\!\!\!\!\!\!\!\!\!\!\!_{\scriptscriptstyle -#3}$}
\def\errorex#1 #2 #3 #4{$#1^{\scriptscriptstyle
+#2}\!\!\!\!\!\!\!\!\!\!\!_{\scriptscriptstyle -#3} \times 10^{#4}$}

\def\farcs{\hbox{$.\!\!{''}$}}

\def\gtwid{\mathrel{\raise.3ex\hbox{$>$\kern-.75em\lower1ex\hbox{$\sim$}}}}
\def\ltwid{\mathrel{\raise.3ex\hbox{$<$\kern-.75em\lower1ex\hbox{$\sim$}}}}

%% Inputs:
\input epsf

%%%%%%%%%%%%%%%%%%%%%%%%%%%%%%%%%%%%%%%%%%%%%%%%%%%%%%%%%%%%%%%%%%%%%%%%%%%%%%%

\centerline{X-RAYS FROM NGC 3256: HIGH-ENERGY EMISSION IN STARBURST GALAXIES}
\smallskip
\centerline{AND THEIR CONTRIBUTION TO THE COSMIC X-RAY BACKGROUND}

\bigskip\bigskip
\centerline{\apj Edward C.\ Moran$^{1,2}$}
\vfootnote{$^1$} {Chandra Fellow.}
\vfootnote{$^2$} {\baselineskip 10pt Participating Guest, Institute of
Geophysics and Planetary Physics, Lawrence Livermore National Laboratory,
Livermore, CA.}
\vfootnote{$^3$} {\baselineskip 10pt Visiting observer at Cerro Tololo
Inter-American Observatory of the National Optical Astronomy Observatories,
operated by Aura, Inc.\ under contract with the National Science Foundation.}
\vfootnote{$^4$} {Also Institute of Astronomy, University of Cambridge,
Cambridge, UK.}

\medskip
\centerline{Department of Astronomy, University of California}
\centerline{Berkeley, CA~94720-3411}

\bigskip\medskip
\centerline{\apj Matthew D.\ Lehnert$^{2,3}$}

\medskip
\centerline{Leiden Observatory}
\centerline{Postbus 9513, 2300RA, Leiden, The Netherlands}

\medskip
\centerline{\apj and}

\medskip
\centerline{\apj David J.\ Helfand{$^4$}}

\medskip
\centerline{Department of Astronomy, Columbia University}
\centerline{New York, NY~10027}

\bigskip\bigskip
\centerline{\apj ABSTRACT}
\bigskip
The infrared-luminous galaxy NGC~3256 is a classic example of a merger-induced
nuclear starburst system.  We find here that it is the most X-ray--luminous
star-forming galaxy yet detected ($L_{\rm 0.5-10\> keV} =$ \EE {1.6} {42}).
Long-slit optical spectroscopy and a deep, high-resolution {\sl ROSAT\/}
X-ray image show that the starburst is driving a ``superwind'' which accounts
for $\sim$~20\% of the observed soft X-ray emission.  Analysis of X-ray
spectral data from {\sl ASCA\/} indicates this gas has a characteristic
temperature of $kT \approx 0.3$ keV.  Our model for the broadband X-ray
emission of NGC~3256 contains two additional components: a warm thermal plasma
($kT \approx 0.8$ keV) associated with the central starburst, and a hard
power-law component with an energy index of $\alpha_{\rm X} \approx 0.7$.
We discuss the energy budget for the two thermal plasmas and find that the
input of mechanical energy from the starburst is more than sufficient to
sustain the observed level of emission.  We also examine possible origins
for the power-law component, concluding that neither a buried AGN nor the
expected population of high-mass X-ray binaries can account for this emission.
Inverse-Compton scattering, involving the galaxy's copious flux of infrared
photons and the relativistic electrons produced by supernovae, is likely to
make a substantial contribution to the hard X-ray flux.  Such a model is
consistent with the observed radio and IR fluxes and the radio and X-ray
spectral indices.  We explore the role of X-ray--luminous starbursts in the
production of the cosmic X-ray background radiation.  The number counts and
spectral index distribution of the faint radio source population, thought to
be dominated by star-forming galaxies, suggest that a significant fraction of
the hard X-ray background could arise from starbursts at moderate redshift.

\medskip\noindent
{\sl Subject headings:} diffuse radiation --- galaxies: individual (NGC~3256) --- galaxies: starburst --- X-rays: galaxies

\vfill
\break
\sec {1.} {Introduction}

\fpsec
The spectacular nearby$^5$ southern galaxy NGC~3256, with a highly disturbed
central region and bright, extended tidal tails, is an excellent example of
a merging system.  Based on its extreme brightness in the near infrared
and strong narrow optical emission-line spectrum, NGC~3256 is considered to be
a ``super starburst'' galaxy (Joseph \& Wright 1985), in the midst of an
especially vigorous episode of star formation.  The starburst and merger
events are expected, ultimately, to render NGC~3256 a gas-depleted elliptical
galaxy (Graham \et 1984).
\vfootnote{$^5$}{$D = 56$ Mpc for $z = 0.0094$ and a Hubble constant of
$H_0 = 50$ km s$^{-1}$ Mpc$^{-1}$, which we use throughout this paper.}

NGC~3256 is seemingly a luminous X-ray source as well.  It was one of 20
``normal'' galaxies from the {\sl ROSAT\/} All-Sky Survey (RASS) reported to
have Seyfert-like X-ray luminosities (Boller \et 1992).  While most of these
X-ray sources did, in fact, prove to be previously unrecognized active galaxies
(Moran, Halpern, \& Helfand 1994), the identification of NGC~3256 as an X-ray
source and its classification as a starburst galaxy remain secure.  The soft
(0.1--2.4 keV) X-ray luminosity of NGC~3256, originally listed as \LX\ =
\EE {2} {42} (Boller \et 1992), would distinguish it as the most luminous
X-ray starburst galaxy currently known (see Fabbiano 1989).

X-ray observations offer valuable insight into the starburst phenomenon and
the cosmological significance of starburst galaxies.  For example, soft X-ray
images have revealed ionized gas in the halos of several starburst
galaxies (\eg Fabbiano 1988), which long-slit optical spectroscopy has
indicated is being driven outward from their nuclear regions (\eg McCarthy,
Heckman, \& van Breugel 1987; Lehnert \& Heckman 1996).
The ultimate fate of this gas has direct
consequences for a number of important outstanding astrophysical problems,
such as the evolution of galaxies and the heating and enrichment of the
intergalactic medium (Heckman 1998).  X-ray--luminous starburst galaxies may
also represent a significant component of the cosmic X-ray background (XRB;
Bookbinder \et 1980; Stewart \et 1982; Griffiths \& Padovani 1990), although to
do so, they would have to exhibit very flat spectra in the hard X-ray band in
order to explain the shape of the XRB spectrum (Fabian \& Barcons 1992).

NGC~3256, because of its apparently exceptional X-ray luminosity, offers a
unique opportunity to investigate these issues.  Furthermore, with an infrared
luminosity of $\sim$~$6 \times 10^{11}\; L_{\odot}$ (Sargent, Sanders,
\& Phillips 1989), NGC~3256 is nearly an ``ultraluminous'' infrared
galaxy.  Hard X-ray observations of this object could also be used to search
for a buried active nucleus and, thus, to comment on the proposition that
ultraluminous infrared galaxies are linked to the origin of quasars (Sanders
\et 1988).  In this paper, we explore both the soft and hard X-ray properties
of NGC~3256 using observations obtained with the {\sl ROSAT\/} and {\sl ASCA\/}
satellites.  In combination with spatially resolved optical spectroscopy and
the results of published radio and infrared studies, these data shed new light
on the nature of the prodigious burst of star formation in NGC~3256.

In \S~2, we present optical evidence for a galactic ``superwind'' driven by
the starburst activity in NCG~3256.  We then present the X-ray imaging and
spectroscopic data we have collected on this galaxy (\S~3).  We interpret
the X-ray emission as arising from three distinct components (\S~4): thermal
emission from both the superwind and supernova-heated gas in the galactic
disk, and a nonthermal component, which is likely to arise from X-ray binary
star systems and inverse-Compton scattered radiation in the nuclear region
of the galaxy.  We close with a discussion of the implications of our analysis
for the origin of the cosmic X-ray background.

\sec {2.} {Optical Spectroscopy of NGC~3256}

\fpsec
Optical spectroscopy provides valuable information regarding the properties
of the ionized gas in galaxies, such as its location, velocity, density,
pressure, ionization state, and heavy-element abundances.  Long-slit spectra
of NGC~3256 were acquired on the night of 1993 March 22 (UT) with the
CTIO~4~m telescope in combination with the R-C Spectrograph.  The spectra,
which have a resolution of 5.6~\AA\ (FWHM) and cover a wavelength range of
4760--6800~\AA , were obtained using a 2\farcs5 slit oriented at position
angles (PAs) of 65$^\circ$ and 155$^\circ$ (east of north).  The exposure
times for the spectra were 900~s (PA = 65$^\circ$) and 1200~s (PA =
155$^\circ$).  Flux calibration was performed using exposures of the stars
L745-46A, LTT~2415, and LTT~7379 taken at the beginning and end of the night
through a 10$''$ slit.  The spectra were summed into 3-column (2\farcs4)
bins in the spatial direction prior to extraction.  For each extracted
spectrum, we measured the fluxes and velocity widths of the following
emission lines: \Hb , \OIII\ $\lambda$5007, \NI\ $\lambda$5199,
\OI\ $\lambda$6300, \Ha , \NII\ $\lambda$6583, and
\SII\ $\lambda\lambda$6716,6731.  We corrected the fluxes of \Ha\ and \Hb\
for absorption due to the underlying stellar continuum by assuming this
absorption has an equivalent width of 5~\AA\ at both wavelengths;$^6$ the
resultant \Ha /\Hb\ ratios were then used to estimate the reddening of
each spectrum, under the assumption that \Ha /\Hb\ has an intrinsic value
of 2.86 (appropriate for Case~B recombination and a gas temperature of
$10^4$~K; Osterbrock 1989).  Based on these reddening estimates, we
corrected the measured emission-line fluxes using the standard extinction
curve from Osterbrock (1989).
\vfootnote {$^6$} {The value of 5~\AA\ for the equivalent width of the stellar
absorption features at \Ha\ and \Hb\ was estimated using high-resolution
($\sim$~1~\AA ) spectra of NGC~3256 taken in the \Hb\ region along PA =
155$^\circ$.  Where the $S/N$ ratio was sufficient (\ie within $\sim$~$10''$
of the nucleus), the wings of the \Hb\ absorption line were clearly resolved
in these data.  Fitting the wings yielded absorption-line strengths of
5$\pm$1 \AA.  This is likely to be the largest correction necessary; thus,
the off-nuclear line ratios we derive are, if anything, underestimated.}

\subsec {2.1.} {Emission-Line Evidence for a Starburst-Driven ``Superwind''}

Observations of nearby edge-on starburst galaxies have revealed both
extended optical emission-line regions and plumes of soft X-ray emission
along the minor axes of these objects, indicating the presence of ionized
gas in their halos (\eg McCarthy \et 1987; Fabbiano 1988; Heckman, Armus,
\& Miley 1990; Armus \et 1995; Lehnert \& Heckman 1995, 1996).  The optical
emission lines from the extraplanar gas exhibit (1) signatures of a complex
velocity field (\ie multiple velocity components, velocity shear, and
velocity widths that increase with increasing distance from the nucleus),
and (2) ``shock-like'' flux ratios (\ie enhancements of the low-ionization
\OI , \OII , \NII , and \SII\ lines relative to \Ha ) similar to those
that characterize low-ionization nuclear emission-line regions (LINERs;
Heckman 1980). Together, these properties suggest that the heating and
kinematics of the ionized halo gas are coupled through shocks, consistent
with the interpretation that this gas arises from a galactic-scale
outflow driven by supernovae in the starburst nucleus (a ``superwind'';
Heckman \et 1990).  The approximately face-on orientation of NGC~3256
($i \approx 30^{\circ}$--$40^{\circ}$; Feast \& Robertson 1978; Moorwood
\& Oliva 1994) makes it difficult to locate the extended line-emitting
regions along the line of sight, and, thus, to compare its emission-line
properties directly with those of edge-on starburst galaxies.  However,
if NGC~3256 does possess a superwind, we might expect it to display the
signatures of a wind far from the nucleus, where the emission is less
dominated by the intense UV field of the bright nuclear \HII\ regions (see
Lehnert \& Heckman 1996).

Reddening- and absorption-corrected emission-line flux ratios for NGC~3256
are plotted in Figures~1 and 2 as a function of projected distance from the
nucleus along PA = 65$^\circ$and PA = 155$^\circ$, respectively.   Line
emission is detected up to half an arcminute (8~kpc) away from the nucleus.
Within the central $10''$, the line flux ratios are consistent with those
of \HII\ regions (Veilleux \& Osterbrock 1987): log~\OIII /\Hb\ $\approx
-0.7$, log~\NII /\Ha\ $< -0.3$, log~\OI /\Ha\ $\approx -1.7$, and
log~\SII /\Ha\ $\approx -0.8$.  Outside the nuclear region, however,
these line ratios increase dramatically, becoming more ``LINER-like''
with increasing distance from the nucleus: log~\NII /\Ha\ $\approx -0.2$,
log~\OI /\Ha\ $> -1.0$, and log~\SII /\Ha\ $> -0.3$.  Figure~3 illustrates
the differences between the nuclear and off-nuclear spectra of NGC~3256
(the off-nuclear spectra are discussed further in the next section).
The velocity widths of the strongest emission lines also increase with
increasing distance from the nucleus, as indicated in Figure~4.  In fact,
a comparison of Figures~2 and~4 reveals that LINER-like emission-line flux
ratios are produced in the {\it same\/} regions where the line velocity
widths are greatest.  This correspondence is demonstrated more clearly in
Figure~5, which shows that the \OI /\Ha\ flux ratio is correlated with the
width of the \Ha\ line.

The coupling between the emission-line flux ratios and velocity widths in
NGC~3256 suggests that the ionization and kinematics of the extranuclear
emission-line gas are strongly influenced by shocks.  Thus, in most respects,
the extended ionized gas in NGC~3256 displays the same characteristics as
the wind-driven halo gas observed in edge-on starburst galaxies (cf.\ Lehnert
\& Heckman 1996).  We note that, because of the roughly face-on orientation
of NGC~3256, we cannot rule out the possibility that some of the extended
line emission is contributed by a diffuse ionized medium (DIM) within the
galaxy.  The DIM observed in the Milky Way (Reynolds 1990; Lehnert \& Heckman
1994) and the disks of late-type spiral galaxies (Wang, Heckman, \& Lehnert
1997) also exhibits strong low-ionization emission lines relative to \Ha .
However, the velocity widths, flux ratios, and surface brightness of the
emission lines associated with the DIM in other galaxies (Wang \et 1997)
are much less extreme than the quantities observed for the gas in the
off-nuclear regions of NGC~3256.  Thus, it is unlikely that a DIM in
NGC~3256, if present, dominates the extended line emission.  We conclude,
therefore, that the starburst in NGC~3256 must be driving a superwind.
As we discuss in \S~3.1, this conclusion is entirely consistent with the
galaxy's soft X-ray morphology.

\subsec {2.2.} {Broad Off-Nuclear \Ha\ Emission}

As illustrated in Figure~3, there appears to be a prominent broad \Ha\
emission line in the off-nuclear spectra of NGC~3256.  Surprisingly, the
broad line is present on either side of the nucleus in both sets of spectra
we obtained.  The onset of this feature in our 2\farcs4 $\times$ 2\farcs5
extractions occurs between 7$''$ and 15$''$ from the nucleus.  In these
spectra, the broad \Ha\ wings have a full-width at zero-intensity (FWZI)
of $\sim (4 - 6) \times 10^3$ km s$^{-1}$.  The line has a FWZI of
$\sim$~6 $\times 10^3$ km s$^{-1}$ in the larger 8$''$ $\times$ 2\farcs5
extractions displayed in Figure~3, and does not appear to be shifted
significantly relative to the narrow component of \Ha .

Because of the broad \Ha\ line, the off-nuclear spectra bear a remarkable
resemblance to the spectra of active galactic nuclei.  However, since
this emission is detected in {\it all\/} of the extended emission-line
regions we examined, the possibility that we are observing AGN light directly
is immediately ruled out.  A scenario in which NGC~3256 contains an
obscured AGN whose broad \Ha\ emission is scattered in our direction by an
off-nuclear ``mirror'' is equally implausible: the combined effects of
(1) the large distance between the nucleus and the scattering medium, (2) the
considerable extinction the AGN light would be likely to suffer en route to
this medium, and (3) the low expected scattering efficiency, suggest that a
buried AGN, if present, would have to be intrinsically very luminous.
Although such an object could be hidden from our optical view, it would be
readily detectable in the radio, mid-infrared, and hard X-ray bands; as
we discuss in \S~4.3.1, there is no evidence at any of these wavelengths
for a luminous, obscured AGN in NGC~3256.

The broad off-nuclear \Ha\ line must, therefore, be related to the
star-formation or merger activity in NGC~3256.  Unfortunately, examples
of starburst galaxies with similar features are scarce.  For instance, in
the extensive survey of edge-on starburst galaxies conducted by Lehnert
\& Heckman (1995), none of the $\sim$~50 galaxies presented exhibits this
characteristic.  Although weak, broad \Ha\ emission with FWZI $\approx$
2000--3000 km~s$^{-1}$ has been observed in a few star-forming galaxies,
the broad lines have been attributed to highly localized phenomena, such
as compact super-star clusters (\eg Heckman \et 1995) or knots of Wolf-Rayet
stars (\eg Sargent \& Filippenko 1991).  Once again, neither of these
explanations seems viable for NGC~3256 since the broad \Ha\ line is observed
throughout the extranuclear emission-line regions.  For this reason, we
suspect that the extended broad \Ha\ emission in NGC~3256 is somehow
associated with the galaxy's starburst-driven wind (cf.\ Tenorio-Tagle \et
1997).  It is possible that the roughly face-on orientation of NGC~3256
has made the detection of this intrinsically faint feature easier than it
would be if the system were edge-on.  Alternatively, the line may be
enhanced in this object because of the ongoing merger activity, which is
likely to have disturbed the gas in the galaxy considerably (\eg Mihos \&
Hernquist 1996).  Comparable data on the extended emission-line properties
of other face-on, merging starburst galaxies is very limited, making it
difficult to assess the relative importance of these factors.

\subsec {2.3.} {Chemical Abundance of the Nuclear Emission-Line Gas}

We have applied the empirical method of Edmunds \& Pagel (1984) to constrain
the heavy-element abundances of the nuclear emission-line gas, which will
assist our modelling of the X-ray spectrum in \S~3.2.  This method
employs reddening-corrected \OIII /\Hb\ and \OIII /\NII\ ratios and assumes
that the excitation processes in the nuclear region of NGC~3256 are similar
to those in the \HII\ regions of nearby galaxies.  As discussed above, this
appears to be the case.  From our dereddened nuclear spectra of NGC~3256, we
find log~\OIII /\Hb\ = $-0.53$ and log~\OIII /\NII\ = $-0.92$.  These ratios
imply an oxygen abundance of 9.1$\pm$0.2 (on the scale 12 + log~[O/H]), which
is approximately 0.2~dex above the solar value.  The uncertainty in
this estimate includes only the uncertainty in the calibration of Edmunds
\& Pagel (1984) and does not reflect possible errors in our measurements of
the line ratios. The latter, however, are relatively small, and we conclude
that metal abundances in the ionized gas in the nucleus of NGC~3256 are
approximately solar.  In \S~4.2.3 we present analysis which suggests that the
abundances of the extended ionized gas in NGC~3256 are similar to those
in the nuclear region.

\sec {3.} {X-Ray Observations of NGC~3256}

\fpsec
NGC~3256 was the target of a 3.4 ks pointed observation with the {\sl ROSAT\/}
Position Sensitive Proportional Counter (PSPC) on 1992 January 14.  We
acquired these data from the HEASARC archive at NASA Goddard Space Flight
Center.  Source counts were collected within a circular region $1'$ in
radius.  Subtraction of the background, which was measured in an annulus
with inner and outer radii of $3'$ and $9'$ centered on the galaxy, resulted
in the detection of 227 source counts in the 0.1--2.4 keV band.  We rebinned
the PSPC spectrum so that it had at least 20 counts per energy channel.

The X-ray source in NGC~3256 appears to be marginally resolved in the PSPC
image, which has an angular resolution of $\sim 30''$.  To obtain a more
accurate determination of the spatial extent of the X-ray emission,
we observed the galaxy with the High Resolution Imager (HRI) on board
{\sl ROSAT\/} for 54.0 ks on 1995 December 10.  The HRI provides $\sim 5''$
resolution, also in the 0.1--2.4 keV band.  A total of 1412 counts above the
background were detected within a $1'$ radius region centered on the galaxy.

We also observed NGC~3256 in the 0.5--10 keV range with the {\sl ASCA\/}
satellite (Tanaka, Inoue, \& Holt 1994) on 1993 December 6.  Total exposures
of 33.2 and 32.4 ks were obtained with {\sl ASCA}'s Solid-state Imaging
Spectrometers SIS0 and SIS1, respectively, which were operated in 1-CCD mode.
An exposure of 36.1 ks was achieved with each of the two Gas Imaging
Spectrometers (GIS2 and GIS3) on board {\sl ASCA}.  Periods of high background
were filtered from the photon event files for each instrument following the
guidelines described in The ABC Guide to {\sl ASCA\/} Data Reduction (Day \et
1995).  Source counts were collected within circular regions  $3'$ (SIS) and
$6'$ (GIS) in radius centered on the galaxy.  The SIS background was measured
in source-free areas on the chip.  In the GIS images, background was measured
in a region with twice the area of the source region located approximately
the same distance off-axis as NGC~3256.
The background-subtracted spectra obtained with the SIS0, SIS1, GIS2, and
GIS3 instruments contain 1116, 1039, 786, and 1000 counts, respectively.
For spectral modeling (\S~3.2), we combined the two SIS spectra and the
two GIS spectra using the FTOOLS task ``addascaspec.'' The resultant spectra
were rebinned to have a minimum of 40 (SIS) or 80 (GIS) counts (source plus
background) per energy channel.

\subsec {3.1.} {Extended Soft X-Ray Emission}

An important key to the nature of the X-ray emission from NGC~3256 is its
spatial extent.  With an angular resolution of $\sim 5''$, the {\sl ROSAT\/}
HRI is best suited to provide this information.  Unfortunately, NGC~3256 is
sufficiently distant that features in the galaxy smaller than $\sim$~1 kpc
in size are unresolved with the HRI.  But as the HRI image displayed in
Figure~6 indicates, the soft X-ray flux from NGC~3256 is significantly more
extended than this.  Although confined to the main body of the galaxy, the
soft X-ray emission spans $45''$ in diameter, or $\sim$~12 kpc.  The X-ray
flux peaks near the optical nucleus at
$\alpha$(2000) = $10^{\rm h} 27^{\rm m} 51.\!\!^{\rm s}$4,
$\delta$(2000) = $-43^{\circ} 54' 15''$,
but a dominant central source, which we would expect if an active nucleus
were responsible for most of the emission, is not present.  In the
full-resolution HRI image ($1''$ pixels, unsmoothed), just 110 counts fall
within a $5'' \times 5''$ detection cell centered on the X-ray peak; as
half the power of the on-axis HRI point-response function is contained in 
a cell of this size (David \et 1997), we derive a firm upper limit of 16\%
for the contribution of a nuclear point source to the HRI counts.  The soft
X-ray half-light radius is about $11''$.

Soft X-ray emission extended over several kiloparsecs is a common property of
starburst galaxies driving hot, outflowing winds, such as in the prototypical
starburst M82 (Moran \& Lehnert 1997, and references therein; see also
Strickland, Ponman, \& Stevens 1997), NGC~253 (Fabbiano 1988), NGC~1569
(Heckman \et 1995), NGC~1808 (Dahlem, Hartner, \& Junkes 1994), NGC~2146
(Armus \et 1995), NGC~3628 (Dahlem \et 1996), NGC~4449 (Della Ceca, Griffiths,
\& Heckman 1997), and Arp~220 (Heckman \et 1996).  As discussed in \S~2.1,
there is ample optical evidence that the starburst in NGC~3256 is also driving
a wind, and much of the extended HRI flux we observe is likely to be associated
with this hot gas.  The X-ray spectrum of NGC~3256, presented in the next
section, provides further details regarding the origin and nature of this
emission.

\subsec {3.2.} {Broadband X-Ray Spectroscopy}

A comprehensive understanding of the X-ray properties of NGC~3256 is only
possible through the analysis of its emission over a wide bandpass, which
our {\sl ASCA\/} observation in the 0.5--10 keV range provides.  Although
{\sl ROSAT\/} is sensitive down to 0.1 keV, very few counts were detected
below 0.5 keV in the short PSPC exposure because of the sizable Galactic
column density in the direction of NGC~3256 ($N_{\rm H} = 9.6 \times 10^{20}$
cm$^{-2}$).  The low $S/N$ ratio PSPC data do not augment the {\sl ASCA\/}
spectra and are, therefore, omitted from our broadband analysis, although
the measured count rates for both the PSPC and HRI detections are consistent
with the model we derive.

We fitted the SIS and GIS spectra of NGC~3256 simultaneously, employing a
variety of models; the results are summarized in Table 1.  The extended nature
of the soft X-ray emission and the optical evidence for a starburst-driven
wind makes a Raymond-Smith (R-S) plasma an obvious model with which to fit
these spectra.  Furthermore, as we demonstrate below, strong emission lines
are plainly visible in the {\sl ASCA\/} spectra, confirming the presence of
a hot, optically thin gas.  But a single R-S component with solar
heavy-element abundances (\S~2.3; see also \S 4.2.3) provides a very poor
fit to the data ($\chi^2$ = 207 for 82 degrees of freedom), so we must
consider more complex models.

As Table 1 indicates, two-component models involving a ``warm'' R-S plasma and
a harder R-S or power law (PL; $F_E \propto E^{-\Gamma}$) component provide
much better fits to the {\sl ASCA\/} spectra of NGC~3256.  In particular, the
fits in which the absorption column densities applied to the two components
are independent yield values of $\chi_{\nu}^2$ (= $\chi^2 / \nu$, where $\nu$
is the number of degrees of freedom) of roughly unity.  In these models, the
characteristic temperature of the warm component is $kT \approx 0.7$ keV; the
hard component has a characteristic temperature of $kT \approx 12$ keV or a
photon index $\Gamma \approx 1.7$.  The {\sl ASCA\/} spectra of NGC~3256,
fitted with the (PL + R-S) model,$^7$\vfootnote{$^7$} {It is unlikely that
the hard X-ray emission of starburst galaxies is dominated by extremely
hot gas (see Suchkov \et 1994); moreover, {\sl ASCA\/} is not capable of
measuring plasma temperatures that lie outside its energy range.  Thus, we
feel the (PL + R-S) form of the two-component model should be preferred.}
are displayed in Figure~7$a$.  Note that the hard component dominates the
X-ray flux below 0.7 keV as well as above $\sim$~1.5 keV.  A nearly identical
circumstance was encountered when a (PL + R-S) model was applied to the
broadband X-ray spectrum of the starburst galaxy NGC 4449 (Della Ceca \et
1997).  However, the dissimilarity of the very soft and hard X-ray
morphologies of that galaxy, which should be the same if a single component
dominates in each band, provided sufficient grounds for the rejection of
two-component models for that object.  In the case of NGC~3256, the specific
absorption scenario suggested in the best two-component models also calls
such models into question.  Despite the significant Galactic column density
toward NGC~3256 ($\sim 10^{21}$ cm$^{-2}$), the best-fit absorption column
for the hard component is zero, whereas the absorption derived for the warm
R-S component is seven to eight times {\it higher\/} than the Galactic value.
Although the formal 90\% confidence ranges for both column densities are
large (see Table~1), they do not overlap, implying that, in these models,
the warm R-S component must be more absorbed.  This is the reverse of the
situation found for other starburst galaxies with sufficiently high $S/N$
ratio broadband X-ray spectra (\eg Awaki \et 1996; Moran \& Lehnert 1997):
the softest thermal emission is more spatially extended and less absorbed
than the hard X-ray component.  For the absorption of the hard X-ray source
in NGC~3256 to be comparable to or greater than that of the warm thermal
component would require the existence of an additional source of soft X-ray
emission in the galaxy.

Unfortunately, the statistical quality of the {\sl ASCA\/} spectra of NGC~3256
is barely adequate for the application of more detailed models.  Nonetheless,
it is becoming clear that the broadband X-ray spectra of starburst galaxies
are, in fact, quite complex.  For example, in our {\sl ROSAT\/} and {\sl 
ASCA\/} study of the nearby starburst M82 (Moran \& Lehnert 1997), we showed
that at least three spectral components are required to fit its high $S/N$
ratio 0.1--10 keV spectrum.  The M82 model consisted of a hard $\Gamma = 1.7$
power law, a warm $kT = 0.6$ keV R-S component, and a soft $kT = 0.3$ keV
R-S component.  We found that while the hard and warm components are
heavily absorbed ($N_{\rm H} \approx 10^{22}$ cm$^{-2}$), the soft R-S
component is not absorbed above the Galactic column.  A nearly identical
model was determined for the {\sl ASCA\/} spectrum of the starburst galaxy
NGC 1808 by Awaki \et (1996).  Similarly, a three-component model was
favored in the analysis of the broadband X-ray spectrum of NGC 4449
(Della Ceca \et 1997).  

Motivated by these precedents and the difficulties
associated with our two-component models, we have applied a three-component
model to the X-ray spectrum of NGC~3256, assuming that the absorption column
density for the hard PL and warm R-S components is the same, and fixing the
absorption column for the soft R-S component at the Galactic value.  As
indicated in Table~1, this model provides a slightly better fit to the
{\sl ASCA\/} spectra than the two-component models do (the improvement is
significant at the 82\% confidence level).  The parameters derived
for the three spectral components are, in some instances, poorly constrained,
but their best-fit values are very similar to those we obtained for M82: for
the hard power-law component, $\Gamma = 1.7$, for the warm thermal component,
$kT = 0.8$ keV, and for the soft thermal component, $kT = 0.3$ keV.  The
column density for the hard and warm components is $N_{\rm H} = 8 \times
10^{21}$ cm$^{-2}$.  As a free parameter, the column density for the soft
component remains near the Galactic value at $1.0 \times 10^{21}$ cm$^{-2}$.
The three-component fit to the {\sl ASCA\/} spectra is displayed in
Figure~7$b$.

On statistical grounds, the two- and three-component models described above
both provide acceptable fits to the {\sl ASCA\/} spectrum of NGC~3256.  As in
the studies of M82 (Moran \& Lehnert 1997) and NGC 4449 (Della Ceca \et 1997),
spatially resolved spectral information would help us to eliminate one of the
possibilities.  Unfortunately, such information will not be available for NGC
3256 until the {\sl Chandra X-ray Observatory\/} is deployed.  But, in
addition to our circumstantial objections to aspects of the two-component
model, the success of the three-component model and its detailed similarity
to the results obtained for starburst galaxies with similar attributes gives
us confidence that it provides the more accurate description of the X-ray
properties of NGC~3256.  Therefore, we adopt that model for the remainder of
this paper.  It is important to note, however, that this choice does not
significantly impact our assessment of the origin of the X-ray emission from
NGC~3256.  For example, the slope and normalization of the power law in the
two- and three-component models are very similar; thus, our conclusions about
the origin and intrinsic luminosity of this emission would be the same in
either scenario.  Likewise, the {\it total\/} intrinsic luminosity of the
thermal emission from the galaxy, computed below in \S~4.1, differs by less
than 25\% in the two- and three-component models.

\sec {4.} {Interpreting the X-Ray Spectrum of NGC~3256}

\fpsec
Because our models for the broadband X-ray emission of NGC~3256 and M82 are
so similar (cf.~Moran \& Lehnert 1997), we interpret their X-ray spectra in
the same manner.  We associate the warm and soft thermal components in the
spectrum of NGC~3256 with the nuclear star formation and the starburst-driven
wind, respectively.  In \S~4.2, we examine the properties of the hot gas and
its relationship to the burst of star formation in NGC~3256.  The origin of
hard X-rays in the galaxy is less clear, and we consider several possibilities
for the source of this emission in \S~4.3.  But first, now that we have an
accurate model of its broadband X-ray spectrum, we reevaluate the luminosity
of NGC~3256 in the X-ray~band.

\subsec {4.1.} {The X-Ray Luminosity of NGC~3256}

The soft X-ray luminosities of starburst galaxies span several orders of
magnitude up to a maximum of a few times \E {41} (Fabbiano 1989).  Thus,
the 0.1--2.4 keV luminosity of \EE {2} {42}, reported from the RASS by Boller
\et (1992), would appear to distinguish NGC~3256 as the most luminous X-ray
starburst currently known.  Although the RASS count rate of 0.113 counts
s$^{-1}$ for NGC~3256 listed in Boller \et has been recently revised to 0.061
counts s$^{-1}$ in the RASS Bright Source Catalogue (Voges \et 1996), NGC~3256
remains, nonetheless, a bright X-ray source.  But with an effective exposure
of just 334~s (Voges \et 1996), the RASS detection of NGC~3256 consists of
only $\sim$~20 counts---too few to provide detailed spectral information,
a key ingredient for the accurate determination of the energy flux.
The considerably more sensitive pointed X-ray observations described above
provide this necessary spectral information.

Despite the fact that our PSPC spectrum of NGC~3256 contains just $\sim$~230
counts, it does allow us to make a direct comparison to the RASS measurement
since it was obtained with the same type of instrument used to conduct the
RASS.  The count rate derived from the pointed PSPC observation is
0.067 counts s$^{-1}$, in agreement with the new RASS count rate
reported by Voges \et (1996).  However, the spectral model used by Boller
\et (1992) to estimate the X-ray flux of NGC~3256---a power law with a
photon index of 2.3---does not provide an acceptable fit to the PSPC spectrum
($\chi^2$ = 33.5 for 10 degrees of freedom).  A model consisting of a $kT
\approx$ 1~keV Raymond-Smith thermal plasma, absorbed by the Galactic neutral
hydrogen column density in the direction of NGC~3256, does a far superior
job ($\chi^2$ = 5.5 for 9 degrees of freedom).  Of course, this is not the
correct form of NGC~3256's X-ray spectrum, but it describes the data in the
0.1--2.4 keV band and thus allows us to determine the flux in that band
accurately.  The fit with the Raymond-Smith model indicates an observed soft
X-ray flux of \FF {6.4} {-13} for NGC~3256.  This translates to an unabsorbed
luminosity of \EE {4.2} {41}, a factor of five lower than the luminosity
quoted by Boller \et (1992).  

However, our {\sl ASCA\/} analysis has revealed the existence of significant
internal absorption in NGC~3256, well in excess of the Galactic column density.
The true soft X-ray luminosity of NGC~3256 should, therefore, be greater than
that which we derived from the fit to the PSPC spectrum.  In Table~2, we list
the observed flux and intrinsic (unabsorbed) luminosity in various X-ray
energy bands for each of the components in the three-component model.  As
expected, a higher value is obtained for the total intrinsic luminosity of
NGC~3256 when the internal absorption is taken into account.  Thus, we confirm
that NGC~3256 is one of the most luminous X-ray starbursts known: in the
0.1--2.4 keV band, \LX\ = \EE {1.9} {42}.  (By a remarkable coincidence,
this luminosity is nearly identical to the luminosity computed by Boller
et al., which was determined using the wrong count rate and an
inappropriate model.)  In the following sections, we investigate in detail
the origin of NGC~3256's copious X-ray emission.

\subsec {4.2.} {Nature of the Thermal X-ray Emission}

We have modeled the soft X-ray emission of NGC~3256 as two thermal plasmas,
one with $kT \approx 0.3$~keV that is not absorbed above the Galactic column,
and one with $kT \approx
0.8$~keV that is moderately absorbed ($N_{\rm H} \approx 10^{22}$~cm$^{-2}$).
Conceptually, we associate the cooler of the two components with the spatially
extended soft X-ray emission (seen in the {\sl ROSAT\/} HRI image) from the
hot, outflowing wind described in \S~2.1.  The absorption and strong emission
lines associated with the warm $kT$ = 0.8 keV plasma suggest that this
component arises from supernova remnants and supernova-heated gas within
a much smaller volume near the starburst nucleus.  Because of the limitations
of the X-ray instrumentation at our disposal and the modest signal to noise
ratio of our spectra of NGC~3256, the true nature of the thermally emitting
gas in the galaxy may actually be more complex.  Nonetheless, the
characteristics of the hot gas in NGC~3256 are intimately linked to the star
formation in the galaxy, and it is instructive to examine whether the
physical parameters we infer for the thermal X-ray--emitting gas are
consistent with other observed and predicted properties of the starburst.

\subsubsec {4.2.1.} {Physical Parameters of the ``Soft'' X-ray--Emitting Gas}

The normalization of the ``soft'' 0.3~keV thermal plasma in the three-component
model listed in Table~1 implies a value for the emission integral $EI = \int
{n_e}\!\! ^2\, dV$ of $4.1 \times 10^{63}$~cm$^{-3}$.  Assuming that the
emission associated with this plasma is uniformly distributed over a spherical
region 6~kpc in radius (corresponding to the extent of the soft X-ray emission
we observe), we derive a total volume for the emitting
region of $V = 2.6 \times 10^{67}$ cm$^{3}$.  In terms of a volume-filling
factor $f_{\rm s}$ for this gas, we compute the following physical parameters
for the soft thermal plasma: the gas density $n_e = (EI / Vf_{\rm s})^{1/2}
= 1.3 \times 10^{-2}\, {f_{\rm s}}\! ^{-1/2}$ cm$^{-3}$, the gas pressure
$P_{\rm X} \approx 2 n_e kT = 1.2 \times 10^{-11}\, {f_{\rm s}}\! ^{-1/2}$
dyne~cm$^{-2}$, the mass $M_{\rm X} \approx n_e m_{\rm p} V f_{\rm s} =
2.7 \times 10^8\, {f_{\rm s}}\! ^{1/2}$~\Msun, and the total thermal energy
$E_{\rm X} \approx 3 n_e kT V f_{\rm s} = 4.6 \times 10^{56}\, {f_{\rm s}}\!
^{1/2}$ ergs.  Using the emissivity of a 0.3~keV gas with solar abundances
$\Lambda = 3.3 \times 10^{-23}$ ergs~cm$^3$~s$^{-1}$ (Sutherland \& Dopita
1993), we calculate a radiative cooling time $t_{\rm cool} \approx 3kT /
(\Lambda n_e) = 1.1 \times 10^8\, {f_{\rm s}}\! ^{1/2}$~yr.

Heckman \et (1990) have measured the density and pressure of the optical
emission-line gas in NGC~3256 (via the \SII\ $\lambda 6716$/ \SII\ $\lambda 
6731$ ratio) as a function of distance from the nucleus.  If this gas were
in pressure equilibrium with the X-ray plasma, we would have a means for
estimating the X-ray volume-filling factor $f_{\rm s}$.  However, the total
pressure of the optical line-emitting gas may be dominated not by its thermal
pressure, but by the ram pressure of the outflowing wind.  Thus, equating
the above relation for the pressure of the X-ray plasma $P_{\rm X}$ with the
(total) pressure of the emission-line gas $P_{\rm opt}$ affords, strictly
speaking, a lower bound on $f_{\rm s}$, which is useful nonetheless.  At a
projected distance of 6~kpc from the nucleus, the Heckman \et density profile
indicates $P_{\rm opt} \approx 1.4 \times 10^{-10}$ dyne~cm$^{-2}$, which
implies $f_{\rm s} \ge 0.01$.  Using this limit for $f_{\rm s}$ in the above
expressions, we have $n_e \le 0.13$~cm$^{-3}$, $M_{\rm X} \ge 2.7 \times
10^7$~\Msun, $E_{\rm X} \ge 4.6 \times 10^{55}$~ergs, and $t_{\rm cool} \ge
1.1 \times 10^7$~yr.  The cooling time is comparable to the sound crossing
time of $\sim 2 \times 10^7$ yr expected for the 0.3 keV plasma.

\subsubsec {4.2.2.} {Physical Parameters of the ``Warm'' X-ray--Emitting Gas}

The normalization of the ``warm'' 0.8~keV thermal plasma in the three-component
model yields an emission integral $EI = 4.9 \times 10^{64}$~cm$^{-3}$.
Unfortunately, our X-ray data do not directly constrain the size of the
region occupied by this plasma.  However, we have spectral evidence that the
X-ray emission originates near the nucleus (inferred from the considerable
absorption found for this component in our fit) and that it is associated with
supernova remnants, suggesting that the 0.8~keV gas is distributed over the
volume within which the bulk of the star formation is occurring.  This is
also the region of mass and energy injection, which, based on the pressure
profile of the optical emission-line gas, has an estimated radius of
$\sim$~1.4 kpc (adjusted for the distance we have adopted for NGC~3256;
Heckman \et 1990).  Assuming spherical geometry, we calculate the volume of
the region responsible for the 0.8~keV thermal emission to be $3.4 \times
10^{65}$~cm$^{3}$, which suggests a pressure for the warm X-ray plasma (in
terms of a volume-filling factor $f_{\rm w}$) of $P_{\rm X} = 1.0 \times
10^{-9} {f_{\rm w}}\!\! ^{-1/2}$~dyne~cm$^{-2}$.  As in the previous section,
we can equate this expression for $P_{\rm X}$ with the pressure of the optical
emission-line gas $P_{\rm opt}$ to obtain a lower limit for $f_{\rm w}$.
At a distance of 1.4~kpc from the nucleus of NGC~3256, $P_{\rm opt} \approx
2 \times 10^{-9}$ dyne~cm$^{-2}$ (Heckman \et 1990), which implies $f_{\rm w}
\ge 0.25$.  Thus, for the warm X-ray--emitting gas, we derive the following
physical parameters: $n_e \le 0.8$~cm$^{-3}$, $M_{\rm X} \ge 5.4 \times
10^7$~\Msun, $E_{\rm X} \ge 2.5 \times 10^{56}$~ergs, and, using a value of
$2.7 \times 10^{-23}$ ergs~cm$^3$~s$^{-1}$ for the emissivity of a 0.8~keV
plasma with solar abundances (Sutherland \& Dopita 1993), $t_{\rm cool} \ge
5.9 \times 10^6$~yr.  Again, the cooling time is comparable to the sound
crossing time of $\sim 3 \times 10^6$ yr expected for this component.

\subsubsec {4.2.3.} {Powering the Thermal X-ray Emission}

Based on the preceding analysis, we estimate the overall mass and energy of the
X-ray--emitting gas in NGC~3256 to be $M_{\rm X} \ge 8.1 \times 10^7$~\Msun\
and $E_{\rm X} \ge 3.0 \times 10^{56}$~ergs.  As indicated in Table~2, the
luminosity of this gas in the 0.1--10~keV band is \LX\ = \EE {1.8} {42}.
We can compare these quantities to the results of starburst models (\eg
Leitherer, Robert, \& Drissen 1992; Leitherer \& Heckman 1995), which predict
the {\it total\/} amount of mass and mechanical energy deposited into the
interstellar medium (ISM) by the burst of star formation.  These models are
calculated as a function of the age of the starburst for the following
parameters: the slope $\alpha$ and upper mass limit $M_{\rm upp}$ of the
stellar mass function, and the metallicity of the star-forming region $Z$.
The set of parameters considered by Leitherer \& Heckman (1995) that most
closely matches the conditions in NGC~3256 are: $\alpha = 2.35$, $M_{\rm upp}
= 100$~\Msun\ (Rigopoulou \et 1996), and $Z = Z_{\odot}$ (\S~2.3).  Assuming
that star formation has been approximately constant over a period of
$\sim 2 \times 10^7$~yr (Rigopoulou \et 1996), these parameters imply a
bolometric absolute magnitude $M_{\rm bol} = -20.60$ for a star-formation
rate of 1 \Msun~yr$^{-1}$ (Leitherer \& Heckman 1995), which corresponds to
a bolometric luminosity of $L_{\rm bol} = 1.4 \times 10^{10}$~$L_\odot$
(\Msun~yr$^{-1}$)$^{-1}$.  Since the majority of the bolometric luminosity
of NGC~3256 is emitted in the far infrared, the FIR luminosity of \LF\ =
$6 \times 10^{11}$ $L_{\odot}$ suggests a star-formation rate of $\sim$~40
\Msun~yr$^{-1}$.  For this star-formation rate, the Leitherer \& Heckman
models predict that $M_{\rm total} = 1 \times 10^8$~\Msun\ and $E_{\rm total}
= 7 \times 10^{57}$~ergs have been released into the galaxy's ISM over the
lifetime of the starburst, and that kinetic energy is being injected into
the ISM at a rate of $2 \times 10^{43}$ ergs~s$^{-1}$.  The relevant time
scale for a comparison of the mechanical energy output of the starburst and
the current thermal energy of the X-ray plasmas is the gas outflow time, which
is approximately equivalent to the sound crossing time of the warm X-ray
plasma.  On this time scale (a few Myr), the starburst injects $E_{\rm X}
\approx 10^{57}$ ergs into the ISM.  Thus, in light of the values of
$M_{\rm X}$, $E_{\rm X}$, and \LX\ listed above for the X-ray plasmas, it
would appear that the starburst in NGC~3256 is indeed capable of powering
the observed thermal X-ray emission.

According to the theory of starburst-driven superwinds, the softest X-ray
emission we observe arises from clouds that are shock-heated by the hot,
outflowing wind material, rather than from the wind itself or supernova
ejecta (Suchkov \et 1994, and references therein).  Thus, there should be
a direct relationship between the rate at which stellar winds and supernovae
are injecting energy into the galaxy and the rate at which the ambient
interstellar medium is being shock-heated.  Assuming (1) that mechanical
energy is being deposited into the ISM of NGC~3256 at a rate of $L_{\rm mech}
= $ \EE {2} {43}, (2) that the temperature of the injected wind material is
equivalent to that of the warm thermal component in our spectral model ($kT
= 0.8$ keV), and (3) that energy is conserved between the warm and soft
thermal plasmas, we derive a mass-heating rate of $\dot M_{\rm heat} =
(L_{\rm mech} m_p) / (3 k T) \approx$ 140 \Msun\ yr$^{-1}$.  The Leitherer
\& Heckman models, on the other hand, predict that the starburst injects
just $\sim$~10 \Msun\ yr$^{-1}$ into the galaxy's ISM, which implies
that the wind in NGC~3256 must be heavily ``mass-loaded'' (Suchkov \et 1994,
1996; see also Della Ceca \et 1997).  The temperature of the injected wind
material may in fact be higher than we have assumed it to be.  If so, the
amount of mass loading would be proportionally lower.  Nonetheless, the
likelihood that the wind of NGC~3256 is mass-loaded suggests that
heavy-element abundances in the soft thermal plasma are similar to those
of the ambient interstellar gas in the galaxy's nuclear region, which we
have demonstrated are approximately solar (\S 2.3).  This substantiates our
adoption of solar abudances for this component in our model of the
X-ray spectrum of NGC~3256.

\subsec {4.3.} {Sources of Hard X-Ray Emission in NGC~3256}

Observations in the 2--20 keV band with the {\sl Ginga\/} satellite first
revealed that star-forming galaxies, as a class, are emitters of hard
X-rays (Ohashi \& Tsuru 1992).  {\sl ASCA}~now provides the opportunity to
disentangle the complex X-ray spectra of these objects and to determine
the energetic processes responsible for their high-energy emission.  An
understanding of the hard X-ray properties of starburst galaxies will lead
to a more complete physical description of the starburst phenomenon and will
permit us to evaluate the contribution of these objects to the cosmic X-ray
background.   In this section we consider sources of hard ($> 2$~keV) X-rays
likely to be associated with the starburst in NGC~3256---accretion-powered
binaries and inverse-Compton scattered emission---as well as the possibility
that the galaxy's hard X-ray emission is produced instead by a hidden Seyfert
nucleus.

\subsubsec {4.3.1.} {An Obscured Active Nucleus}

As we demonstrated in \S~4.1, the total X-ray luminosity of NGC~3256 is high
for a starburst galaxy.  In addition, the best-fit slope of the power-law
component we derive (energy index $\alpha_{\rm X} = 0.7$) is similar to the
slope of a typical Seyfert galaxy X-ray spectrum (Nandra \& Pounds 1994).
The amount of
absorption applied to this component in our fit ($N_{\rm H} \approx 10^{22}$
cm$^{-2}$) could be interpreted as evidence that NGC~3256 contains an obscured
low-luminosity Seyfert nucleus.  Based on the IR brightness of NGC~3256 (\LF\
= $6 \times 10^{11}\; L_{\odot}$) and the popularity of an evolutionary
scenario linking ultraluminous IR galaxies to the origin of quasars (Sanders
\et 1988), several authors have addressed the question of whether or not
NGC~3256 harbors a buried AGN.  But observations in the near-infrared
(Moorwood \& Oliva 1994; Kotilainen \et 1996), radio (Norris \& Forbes 1995),
and optical (\S~2) bands have {\it all\/} failed to indicate Seyfert activity
in NGC~3256.  Mid-infrared spectroscopy obtained with the {\sl Infrared Space
Observatory (ISO)\/}, which can penetrate obscuration equivalent to $A_V
\approx 50$~mag, has provided the most compelling evidence to date that
the burst of star formation in NGC 3256---{\it not\/} a buried active
nucleus---powers the galaxy's luminosity.  High-excitation emission lines,
such as [O~{\apj iv}] $25.9\mu$ and [Ne~{\apj v}] $14.3\mu$, which are
indicative of photoionization by a hard, nonstellar continuum, were not
detected in the {\sl ISO\/} spectrum of NGC~3256 (Rigopoulou \et 1996).
Instead, photoionization by hot stars provides a fully adequate explanation
for the IR emission-line properties of the galaxy (Lutz \et 1996).  Given
the absence of any evidence for an AGN in NGC~3256, we are reluctant to invoke
a buried Seyfert nucleus to explain the galaxy's hard X-ray emission.

\subsubsec {4.3.2.} {X-ray Binaries}

The identification of high-mass X-ray binary star systems (HMXBs) in the
Galaxy and Magellanic Clouds has led to the expectation that such sources
contribute significantly to the total X-ray emission of other star-forming
galaxies (Fabbiano 1989).  Since HMXBs have distinctly flat X-ray spectra
below 10~keV ($\bar{\Gamma} \approx 1.2$; White, Swank, \& Holt 1983; Nagase
1989), the energy range and
resolution afforded by {\sl ASCA\/} should permit a sensitive spectral test
of this hypothesis.  Yet despite {\sl ASCA}'s capabilities, the role of HMXBs
in starburst galaxies remains unclear.  While the hard X-ray spectra of some
starbursts may be consistent with those of Galactic HMXBs (Yaqoob \et 1995;
Awaki \et 1996), the spectra of other objects clearly are not (Moran \& Lehnert
1997; Della Ceca \et 1996, 1997; Ptak \et 1997). Unfortunately, the constraints
we have derived here for the slope of the hard component in the X-ray spectrum
of NGC~3256 are insufficient for a meaningful comparison to the spectra of
HMXBs.  Thus, we must consider the contribution of HMXBs to the energy budget
of NGC~3256 instead.  

HMXBs consist of an early-type star and the compact remnant of an early-type
star.  Therefore, the number of O~stars in a star-forming region should
provide an indication of the number of HMXBs present.  If the star-formation
rate is approximately constant, O~star births and deaths will reach equilibrium
in $\sim$~$10^7$ yr, after which time the number of O~stars will remain
constant (Leitherer \& Heckman 1995). Since the duration of the X-ray--emitting
phase of a high-mass binary system is much shorter than the lifetimes of even
the most massive stars (\eg Dalton \& Sarazin 1995), the HMXB population will
also reach equilibrium on this time scale.  Thus, under these conditions, the
number of HMXBs in a star-forming region should be directly proportional to
the number of O~stars present, regardless of the star-formation rate.

To determine the number of O~stars in the nucleus of NGC~3256, we have
employed the method of Vacca (1994), which requires an estimate of $Q$, the
number of ionizing photons (\ie those with $\lambda < 912$~\AA ) emitted
per second by O~stars.  Based on the {\sl ISO\/} observation of NGC~3256
(Rigopoulou \et 1996), Lutz \et (1996) computed a Lyman continuum luminosity
of $2.9 \times 10^{10}$ $L_{\odot}$, which, after adjustment for a distance
of 56 Mpc and their adopted mean photon energy of 15~eV, corresponds to $Q =
1.0 \times 10^{55}$ photons s$^{-1}$.  If we assume solar metallicity for the
star-forming region ($Z \approx Z_{\odot}$; \S~2) and a stellar upper mass
cutoff $M_{\rm upp}$ of 80--100 \Msun\ (Rigopoulou \et 1996), and allow for
a range of possible values of the slope of the initial mass function $\alpha$
= 2.0--2.7, we estimate that there are between $7.0 \times 10^5$ and $1.3
\times 10^6$ O~stars in the starburst nucleus of NGC~3256.  This is consistent
with the value of $1.0 \times 10^6$ O~stars predicted by Leitherer \& Heckman
(1995) for a constant star-formation rate of 40 \Msun\ yr$^{-1}$, $\alpha =
2.35$, $M_{\rm upp} = 100$~\Msun , $Z = Z_{\odot}$, and a starburst age of
$> 10^7$ yr.  As indicated in Table~2, the 2--10 keV luminosity of the
power-law component in the spectrum of NGC~3256 is \EE{1.9} {41}.  Given
the number of O~stars calculated here, we require a specific hard X-ray
luminosity for NGC~3256 of (1.5--2.7) $\times$ \E {35} per O~star.

This value is substantially in excess of that found in the Milky Way, an
appropriate system for comparison given the solar abundance we measure in
NGC~3256. When calculating the specific hard X-ray luminosity per O~star,
it is essential to define with care the X-ray luminosity of each contributing
HMXB system. All such X-ray sources are variable; indeed, the vast majority
of HMXBs by number are the Be star X-ray binaries, which have very small duty
cycles and very low quiescent X-ray luminosities.  Furthermore, catalogued
fluxes for HMXB systems are quoted for a variety of different observing
bands and assumed spectral forms, and often include only the maximum flux
ever observed, or the maximum and minimum recorded fluxes (where the latter
is usually an upper limit) and no duty cycle information (cf.\ Bradt \&
McClintock 1983; van Paradijs 1995). While these data are sufficient for most
applications, they do not allow a determination of the {\it time-averaged\/}
source luminosity required for a calculation of the specific X-ray luminosity
per O~star.

To address this issue, we have carried out an extensive review of the
literature and various data archives, from which we have assembled a
volume-limited sample of nearly 60 HMXBs within 3.5~kpc of the Sun.  Using
the available all-sky-monitor databases, we have determined the mean
2--10 keV luminosity of each source in this list (Helfand \& Moran 1999).
In conjunction with the comprehensive new catalog of OB stars in the solar
neighborhood compiled by C.~Garmany (1998, private communication), this
provides a firm upper limit on the specific luminosity of \EE {4} {34}
per O~star; the best estimate is a factor of two lower.  These values are
consistent with that obtained for the Galaxy as a whole using the HMXB
population syntheses of Portegies Zwart \& Verbunt (1996) and Dalton \&
Sarazin (1995).  Thus, we conclude that less than $\sim$ 20\% of the hard
X-ray emission from NGC~3256 arises from its HMXB population.

\subsubsec {4.3.3.} {Inverse-Compton Scattering}

Inverse-Compton scattering (IC), involving the interaction of infrared
photons with supernova-generated relativistic electrons, was among the first
mechanisms proposed to explain the emission of hard X-rays from M82 (Hargrave
1974).  Although the IC hypothesis is difficult to test, the luminosities of
M82's cospatial mid-infrared and nonthermal radio sources indicate that the
IR photons and relativistic electrons possess energy densities high enough
for the production of a significant X-ray flux through the IC process (Rieke
\et 1980).  Observations of the broadband X-ray spectrum and luminosity of
M82 have pointed to IC emission as an important source of X-rays at energies
above a few keV (Schaaf \et 1989; Moran \& Lehnert 1997).  The nuclear region
of NGC~3256 is also a luminous source of infrared and radio emission.  Given
the shortcomings of other explanations for its hard X-ray flux, the role of
IC scattering in NGC~3256 should be explored as well.

The calculation of the expected IC X-ray luminosity depends on the geometry
adopted for the emitting region, which can only be approximated for an
object as distant as NGC~3256.  Fortunately, some assistance is provided
by recent high-resolution radio (Norris \& Forbes 1995) and near-IR
(Kotilainen \et 1996) images of NGC~3256.  These have revealed the presence
of two resolved nuclei in the galaxy separated by $\sim 5''$.  The 5~GHz
flux densities of the nuclei are nearly identical (34~mJy and 31~mJy),
as are their radio spectral indices ($\alpha_{\rm r} = 0.78$ and 0.86 for
$S_{\nu} \propto \nu^{-\alpha_{\rm r}}$).  For simplicity, therefore, we
will assume that $S_5 = 33$ mJy and $\alpha_{\rm r} = 0.8$ for each nucleus,
and that each is responsible for half of the galaxy's infrared luminosity
of \EE {2.4} {45}.  Assuming further that the nuclei are spherical, their
measured radio sizes (1\farcs2 FWHM; Norris \& Forbes 1995) imply they
have radii of 0\farcs7 (190 pc).

If the relativistic electrons and inverse-Compton seed photons are cospatial
with a uniform magnetic field $B$, the IC luminosity can be expressed as
$L_{\rm IC} = L_{\rm synch} \times (U_{\rm ph} / U_{\rm B})$, where
$L_{\rm synch}$ is the total synchrotron luminosity produced by the electrons,
$U_{\rm ph}$ is the energy density of the seed photon field, and $U_{\rm B}$
is the energy density of the magnetic field.  Integration of the radio
spectrum of NGC~3256 between 10~MHz and 200~GHz yields $L_{\rm synch}$ = \EE
{5.7} {39} for each nucleus.  For a spherical region of radius $r$, the seed
photon energy density is given by $U_{\rm ph} = 3 L_{\rm bol} / 4 \pi r^2 c$,
where $L_{\rm bol}$ is the bolometric luminosity of the region and $c$ is the
speed of light.  In NGC~3256, $L_{\rm bol} \approx$ \LF, implying $U_{\rm ph}$
= $2.8 \times 10^{-8}$ ergs cm$^{-3}$ in each nucleus.  An estimate of
$U_{\rm B}$ can be derived under the assumption that the minimum energy
condition obtains in the synchrotron-emitting plasma.  From equation (2) of
Miley (1980), we calculate a minimum energy magnetic field strength of
$B_{\rm min} = 70\; (1 + k)^{2/7}$~$\mu$G.  Here $k$ is the proton-electron
energy ratio, which, although unmeasured, is thought to lie between 1 and 100.
For this range of $k$, $U_{\rm B}$ (= ${B^2}_{\rm\nts\nts\nts min}/8\pi$) has
a value of $2.9 \times 10^{-10}$ to $2.7 \times 10^{-9}$ ergs~cm$^{-3}$.
The combined IC luminosity of both nuclei, integrated over the entire range
of upscattered energies, is therefore expected to be between \EE {1.2} {41}
and \EE {1.1} {42}, suggesting that IC scattering may indeed account for a
significant fraction of the hard X-ray emission of NGC~3256.

To compute the IC luminosity emitted in the {\sl ASCA\/} band, we employ the
approach described by Tucker (1975).  This method capitalizes on the fact
that electrons possessing a power-law distribution of energies with index
$n$ produce power-law synchrotron and IC spectra with identical energy indices
$\alpha = (n - 1)/2$.  (Note that the hard X-ray and nuclear radio spectra
of NGC~3256 {\it do\/} have approximately the same slopes.)  Assuming the
seed photon field has a blackbody spectrum of temperature $T$, the ratio of
the IC and synchrotron spectral power per unit volume can be expressed as
$$K = 2.47 \times 10^{-19}\; (5.25 \times 10^3)^{\alpha}\;  T^{3 + \alpha}\; 
B^{-(1 + \alpha)}\; b(n)/a(n)\, . \eqno{(1)}$$  The value of $0.8$ we have
adopted here for $\alpha$ implies $n = 2.6$, which corresponds to values of
0.08 and 8.4 for the dimensionless functions $a(n)$ and $b(n)$, respectively.
The observed 60~$\mu$m to 100~$\mu$m {\sl IRAS\/} flux density ratio for 
NGC~3256 ($S_{60}/S_{100} = 0.77$) suggests a dust temperature of $T = 44.4$~K
(assuming a modified blackbody in which the dust emissivity is proportional
to the frequency $\nu$; see Helou \et 1988).  Thus, for the range of
$B_{\rm min}$ derived above, $K$ = 0.13--0.95.

The quantity $K$ can then be used to relate the expected IC flux density
$F_{\rm IC}$ at an X-ray frequency $\nu_{\rm IC}$ to the observed
synchrotron flux density $F_{\rm s}$ at a radio frequency $\nu_{\rm s}$:
$$K = (F_{\rm IC} / F_{\rm s}) \times (\nu_{\rm IC} / \nu_{\rm s})^{\alpha}\, .
\eqno{(2)}$$  Using the total nuclear 5~GHz radio flux density of NGC~3256,
we compute an expected 5~keV IC flux density of $(0.4 - 3.0) \times
10^{-14}$ ergs cm$^{-2}$ s$^{-1}$ keV$^{-1}$.  Comparison to the observed
5~keV flux density of NGC~3256 of $\sim 6.4 \times 10^{-14}$ ergs cm$^{-2}$
s$^{-1}$ keV$^{-1}$ (see the lower panel of Fig.~7b) suggests that IC
scattered emission accounts for 6--47\% of the hard X-ray flux
in the {\sl ASCA\/} bandpass.

Given the value for the binary contribution obtained above ($< 20$\%), it
would appear that our simple estimate yields an IC flux that still falls
short of the total measured hard X-ray flux from NGC~3256.  However, it is
important to bear in mind that the IC calculation is highly idealized: we
have assumed that (1) the geometry of the starburst region is simple, (2)
the electron, photon, and magnetic field energy densities are uniform, and
(3) the minimum energy scenario is valid.  Deviations from these conditions
could have a significant impact on the actual IC luminosity produced.  For
instance, a substantially higher IC flux would result if the ambient magnetic
field strength were slightly below the equipartition value.  In addition,
the spatial distribution of the relativistic electrons and infrared photons
may be knotty, amplifying the local electron and photon energy densities
in some fraction of the star-forming region.  In light of these uncertainties,
it is possible that the role of IC scattering in NGC~3256 may be more
significant than our calculations above would suggest.  High-resolution radio,
infrared, and hard X-ray observations are needed to improve our understanding
of the importance of IC scattering in this and other starburst nuclei.

\sec {5.} {Starburst Galaxies and the X-ray Background}

\fpsec
Perhaps the most pressing unsolved problem in extragalactic X-ray astronomy
is the origin of the cosmic X-ray background (XRB).  It is now known that
the XRB above $\sim$~1~keV is produced by discrete emitters rather than
arising in a truly diffuse medium (Mather \et 1990; Wright \et 1994).  But
despite vigorous attempts, efforts to resolve the XRB completely have not
yet succeeded.  In an extremely deep {\sl ROSAT\/} survey, approximately
70\% of the 1--2 keV XRB was resolved into discrete sources (Hasinger \et
1998).  Follow-up optical spectroscopy has revealed that, as in shallower
soft X-ray surveys, the majority of these sources are high-luminosity active
galactic nuclei (Schmidt \et 1998).  Paradoxically, the broadband X-ray
spectra of such sources are considerably steeper than the spectrum of the
XRB itself (\eg Nandra \& Pounds 1994); thus, the objects that dominate
the XRB in the soft band contribute only a small fraction of the background
at higher X-ray energies (Fabian \& Barcons 1992). It remains unclear (1)
what population is responsible for the as-yet unresolved soft XRB, and (2)
if these sources, once identified, will exhibit the spectral properties
required to account for the overall shape of the XRB spectrum.

Although it is currently fashionable to attribute the balance of the XRB
radiation to a population of intrinsically absorbed Seyfert galaxies (\eg
Madau, Ghisellini, \& Fabian 1994; Comastri \et 1995), it has been proposed
on a number of occasions that starburst galaxies might also make a significant
contribution to the XRB (Bookbinder \et 1980; Stewart \et 1982; Weedman 1987;
Griffiths \& Padovani 1990; Rephaeli \et 1991; David \et 1992).  The emission
of hard X-rays appears to be ubiquitous among starbursts; therefore, their
integrated contribution could rival that of classical AGNs, which, despite
being more
luminous, are considerably more rare.  Previously, it has been assumed that
HMXBs dominate the emission of star-forming galaxies above a few keV (\eg
Bookbinder \et 1980; Griffiths \& Padovani 1990).  In our investigations,
we have concluded that inverse-Compton scattering provides an equally plausible
explanation for the hard X-ray luminosities of both NGC~3256 (\S~4) and M82
(Moran \& Lehnert 1997).  Independent of the emission mechanism, however,
we can use these new measurements of the hard X-ray flux from starbursts
to estimate the contribution such galaxies make to the XRB.

A key aspect of any such calculation is the assumption concerning the
evolution of the starburst galaxy population with cosmic time.  Rather than
attempt to model this evolution, we seek an observational contraint on the
total number of potential contributors.  It has long been established that
there exists a very tight correlation between the far-infrared emission of
(non-AGN) galaxies and their centimetric radio flux densities (Helou, Soifer,
\& Rowan-Robinson 1985).  While a detailed physical explanation for this
correlation is not yet in hand, it is apparent that star-formation is
responsible for generating the bulk of the radiation in both bands---in the
far infrared from stellar UV radiation reprocessed by dust, and in the radio
from \HII\ region free-free emission plus synchrotron radiation generated
by the relativistic electrons accelerated in supernova explosions.  Thus, the
5~GHz radio emission from galaxies lacking a radio-loud AGN can be used
as a surrogate for estimating their star formation activity.

Since current sensitivities in both the far-IR and X-ray bands are
insufficient to probe star forming populations directly at intermediate
and high redshifts, we base our analysis on the well-constrained radio
source counts.  Below a flux density of $\sim$~1 mJy, the radio source
population is associated predominantly with faint, blue (and presumably
star-forming) galaxies, rather than with quasars and early-type galaxies,
which account for the majority of sources at higher flux densities (\eg
Windhorst \et 1985; Kron, Koo, \& Windhorst 1985; Thuan \& Condon 1987;
Condon 1989; Benn \et 1993; Windhorst \et 1995; Richards \et 1998).  The
5~GHz number count--flux relation ($\log N - \log S$) measured by Fomalont
\et (1991) thus provides the means by which we can estimate the integrated
contribution of starburst galaxies to the XRB.  Over a flux density range
of 16~$\mu$Jy to 1~mJy, Fomalont \et found that, for $S$ in units of
$\mu$Jy, $N(> S) = 23.2\; S^{-1.2}$ arcmin$^{-2}$. On the basis of
fluctuation statistics, they showed further that the relation continues
with approximately the same slope down to $\sim$~2~$\mu$Jy.  In differential
form, the microjansky number counts can thus be expressed as $n(S) = -27.8\;
S^{-2.2}$ arcmin$^{-2}$ = $-1.0 \times 10^5\; S^{-2.2}$ deg$^{-2}$.  Provided
we can relate the expected X-ray emission from a galaxy directly to its radio
flux density, the following equation then describes the contribution of
starburst galaxies to the intensity of the XRB: $$I_{\rm X} = \int n(S)\; 
f_{\rm SB}(S)\; F_{\rm X}(S)\; dS \eqno{(3)}$$ where $f_{\rm SB}(S)$ is
the fraction of radio sources with a 5~GHz flux density of $S$ that are
starburst galaxies, and $F_{\rm X}(S)$ is the X-ray flux.

As we have discussed above, IC emission in starburst galaxies provides a
direct connection between their radio and X-ray emission; however, since
it depends in part on the geometry of the star-forming region, it is not
obvious that this emission will scale linearly with the radio flux density.
Alternatively, if HMXBs and/or supernova remnants produce the bulk of the
hard X-ray emission, the link between radio flux density and X-ray output
may be even less direct. Nonetheless, in both starburst galaxies for which
we have a good measurement of the hard X-ray flux, $R_{5,5}$, the ratio of
the 5~keV flux density$^8$\vfootnote{$^8$} {An X-ray energy of 5 keV is
ideal for this calculation: the 5~keV flux densities of starburst galaxies
observed with {\sl ASCA\/} are relatively well determined, and they are
unaffected by modest photoelectric absorption or the soft thermal emission
components which dominate at lower energies.}
to the {\it core\/} 5~GHz flux density, has approximately the same value.
In NGC~3256, the 5~keV flux density is $6.4 \times 10^{-14}$ ergs cm$^{-2}$
s$^{-1}$ keV$^{-1}$ and the core 5~GHz flux density is 65~mJy (Norris \&
Forbes 1995). In M82, the 5~keV flux density is $3.2 \times 10^{-12}$ ergs
cm$^{-2}$ s$^{-1}$ keV$^{-1}$ (Moran \& Lehnert 1997) and the core 5~GHz
radio flux density is 3.4~Jy (Hargrave 1974).  Thus, for both galaxies
$R_{5,5} \approx 10^{-18}$ ergs cm$^{-2}$ s$^{-1}$ keV$^{-1}$ $\mu$Jy$^{-1}$.

In the IC picture, the hard X-ray and GHz radio spectra of starburst galaxies
are expected to have the same slope, so $R_{5,5}$ does not need to be adjusted
for redshift effects (in other scenarios, a $K$-correction may be required).
We do, however, need to account for the fact that only a fraction
$f_{\rm core}$ of the total observed radio flux of a starburst galaxy
is emitted in the nuclear region.  In M82, $f_{\rm core} = 0.85$ (Hargrave
1974), which we will adopt as typical for starburst galaxies.  Thus, for a
given {\it total\/} 5~GHz flux density $S$, we estimate the expected 5~keV
flux density as $F_{\rm 5\; keV} = R_{5,5}\; f_{\rm core}\; S$ ergs
cm$^{-2}$ s$^{-1}$ keV$^{-1}$; equation (3) then becomes $$I_{\rm 5\; keV}
= -1.0 \times 10^5 \int R_{5,5}\; f_{\rm core}\; f_{\rm SB}(S)\; S^{-1.2}\;
dS\; . \eqno{(4)}$$  Assuming, for the time being, that $f_{\rm SB}$ does
not vary with $S$, evaluation of the integral yields $$I_{\rm 5\; keV} =
4.3 \times 10^{-13}\; f_{\rm SB}\; \bigl[{S_1}\!^{-0.2} - {S_2}^{-0.2}\bigr]\;
{\rm ergs}\; {\rm cm}^{-2}\; {\rm s}^{-1}\; {\rm keV}^{-1} \; {\rm deg}^{-2}
\eqno{(5)}$$ where $S_1$ and $S_2$ are, respectively, the lower and upper
limits of integration, in units of $\mu$Jy.

The change in the slope of the radio $\log N - \log S$ relation, below which
radio sources become associated with faint blue galaxies, occurs somewhere
in the neighborhood of one to a few mJy (Fomalont \et 1991, and references
therein).  For the power-law form of this relation used in equation~(4),
sources with fluxes in excess of these values make virtually no contribution
to the integral, so we may take $S_2$ to be infinity.  Our estimate of
$I_{\rm 5\; keV}$ depends, therefore, entirely on the value we adopt for $S_1$.
Windhorst \et (1993) have shown that the radio source counts at 8.3~GHz
must converge below 20~nJy; otherwise, discrete radio sources would distort
the cosmic microwave background spectrum at centimeter wavelengths.  The
corresponding limit for 5~GHz sources, if we apply the mean spectral index
of $-0.35$ measured by Windhorst \et (1993) for sources in the microjansky
regime, is 24~nJy.  Thus, assuming $f_{\rm SB} \approx 1$ (which could well be
appropriate for sub-$\mu$Jy sources) and $S_1 = 0.024\; \mu$Jy, equation~(5)
yields $I_{\rm 5\; keV} = 9.1 \times 10^{-13}$ ergs cm$^{-2}$ s$^{-1}$
keV$^{-1}$ deg$^{-2}$.  Comparing this value to the measured intensity of
the XRB at 5~keV of $2 \times 10^{-12}$ ergs cm$^{-2}$ s$^{-1}$ keV$^{-1}$
deg$^{-2}$ (Gendreau \et 1995), we find that star-forming galaxies may
contribute as much as 45\% of the 5~keV XRB.

Analysis of optical galaxy counts (Windhorst \et 1993) and the far-infrared
background (Haarsma \& Partridge 1998) have suggested that the radio source
population may dwindle at fluxes as high as 0.3--1~$\mu$Jy.  In addition,
optical identifications of 10--1000 $\mu$Jy radio sources indicate that only
$\sim$~70\% to 80\% are likely to be associated with star-forming galaxies
(Windhorst \et 1995; Richards 1998).  Thus, if we conservatively assume that
$S_1 = 1\; \mu$Jy and $f_{\rm SB} = 0.75$, equation (5) yields $I_{\rm 5\; keV}
= 3.2 \times 10^{-13}$ ergs cm$^{-2}$ s$^{-1}$ keV$^{-1}$ deg$^{-2}$, 
equivalent to $\sim$~15\% of the 5~keV XRB.

The Cosmic Infrared Background (CIB), recently measured from an analysis of
the {\sl COBE\/} satellite data (Dwek \et 1998, and references therein),
provides
another constraint on the fraction of the background in any band contributed
by processes associated with star formation.  The energy density of the CIB
in the 125~$\mu$m to 5000~$\mu$m band is $(6.7 \pm 1.7) \times 10^{-15}$ ergs
cm$^{-3}$ (from Dwek \et 1998), while between 5 and 35 keV, the band for
which a new population of flat-spectrum XRB contributors is required, the
observed energy density is $\sim 3.5 \times 10^{-17}$ ergs cm$^{-3}$.  This
implies that only $\sim 10^{-3}$ of the CIB radiation needs to be processed
into hard X-rays in order to make a 25\% contribution to the XRB flux. The
far-infrared--to--X-ray luminosity ratio for NGC 3256 is $\sim 2 \times
10^{-4}$; thus, if all starbursts contribute at this level, the overall
background contribution would be $\sim$~5\%.  It is not difficult to imagine
evolutionary effects (in the IMF slope, the compactness of the nuclear
starbursts, the metallicity, etc.)~that would raise this contribution
considerably.  Furthermore, Steidel \et (1999) have recently shown that star
formation activity in galaxies at $\langle z \rangle \approx 4.1$ is
significant, with
very modest levels of obscuration by dust; such UV/visible emitters would
not contribute significantly to the CIB, but could well contribute to the
XRB an amount comparable to dust-enshrouded star-forming regions.  Since the
optical/UV background light is roughly equal in energy density to the CIB,
the contribution to the XRB by star formation could increase by an additional
factor of $\sim$~2.

A number of issues must be investigated further in order to refine our
estimate of the contribution star-forming galaxies make to the XRB.  First,
we require a better understanding of the relationship between the radio and
hard X-ray properties of starburst galaxies.  A definitive determination of
the origin of their hard X-ray emission would facilitate
this understanding.  Second, the function $f_{\rm SB}(S)$ needs to be
determined through optical identification programs for the faintest known
radio sources (e.g., Richards 1998).  Finally, the nature of the
radio $\log N - \log S$ relation below $S \approx 1$~$\mu$Jy is of crucial
importance, since a major fraction of the XRB contribution of starbursts
could arise from sub-$\mu$Jy sources.

It remains to be seen whether the average X-ray spectra of distant star-forming
galaxies are flat enough to help resolve the XRB spectral paradox---that
all identified constributors have spectra steeper than that of the integrated
background itself.  However, indications from the microjansky radio source
statistics are encouraging in this regard.  Windhorst \et (1993) have
demonstrated that the spectra of microjansky radio sources, most of which
are likely to be star-forming galaxies, flatten to $\bar{\alpha}_{\rm r} =
0.35 \pm 0.15$ at the faintest measured flux densities.  This is consistent
with both the slope of the XRB spectrum of $\alpha_{\rm X} = 0.41 \pm 0.03$
measured by Gendreau \et (1995) and the IC model requirement that the
radio and X-ray spectral slopes be the same.  Such a hypothesis is easily
testable: a 500~ks {\sl Chandra\/} pointing would be able to detect NGC~3256
to $z$ = 2, and will contain dozens of starburst galaxies coincident with
radio sources brighter than $\sim$~10 $\mu$Jy if the IC radiation from
starbursts produces a significant contribution to the hard XRB.

\sec {6.} {Summary}

\fpsec
With a 0.5--10 keV luminosity of \EE {1.6} {42},
NGC~3256 is the most X-ray--luminous starburst galaxy currently known.
Our optical spectroscopy and deep, high-resolution soft X-ray image of
this object indicate that, similar to other nearby galaxies undergoing
intense nuclear bursts of star formation, NGC~3256 has a significant
amount of ionized gas in its halo that appears to possess all the
characteristics of a ``superwind.''  We associate this gas with the
0.3~keV thermal component found in our model for the broadband {\sl ASCA\/}
spectrum of NGC~3256.  A second thermal component in this spectrum, with
$kT = 0.8$~keV, is likely to be associated with supernova-heated gas
confined to the starburst region itself.  Our calculations indicate
that the star formation in NGC~3256 is fully capable of supplying the
mass and energy contained in these hot plasmas.

Hard X-rays are detected from NGC~3256 with energies up to 10~keV.  Contrary
to expectations that such emission in starburst galaxies is produced mainly by
high-mass X-ray binary systems, our comparison of the stellar population in
NGC~3256 with nearby star-forming regions indicates that HMXBs are unable
to account for more than $\sim$~20\% of the galaxy's hard X-ray luminosity.
Despite the detection of broad \Ha\ features in the extended emission-line
regions, we find no evidence for a buried active nucleus in this object.  We
have shown that inverse-Compton scattered emission, involving infrared photons
and relativistic electrons produced as a result of the vigorous star formation
in NGC~3256, is likely to provide a substantial fraction of its high-energy
flux.  The assumption that inverse-Compton scattering is generally responsible
for the majority of the hard X-ray emission of star-forming galaxies explains
the coincidence in spectral slope between faint radio sources and the hard
XRB.  Independent of the nature of the X-ray emission from starbursts, however,
we find that these objects can make a significant contribution to the hard
X-ray background.

\bigskip
We begin by acknowledging the extraordinarily thorough and helpful report of
the referee, Prof.\ T.\ Heckman; incorporation of his input has improved this
paper significantly.  We also thank the staff at CTIO for the generous
allocation of telescope time and for their assistance obtaining the optical
data presented in this paper.  We are indebted to Rob Petre for his
assistance with the planning of the {\sl ASCA\/} observation, and to
Amiel Sternberg for many helpful discussions about starburst galaxies.
Our research has made use of {\sl ROSAT\/} archival data obtained through
the High Energy Astrophysics Science Archive Research Center (HEASARC)
at NASA Goddard Space Flight Center.  This work has been generously
supported by NASA through grants NAG5-2556, NGG5-6035, and NAG5-3556.
ECM acknowledges partial support by NASA through Chandra Fellowship grant
PF8-10004 awarded by the Chandra X-ray Center, which is operated by the
Smithsonian Astrophysical Observatory for NASA under contract NAS8-39073.
The work of MDL was funded through the Dutch Organization for Research
(NWO) and the Dutch Ministry of Education.  Support for ECM and MDL at
IGPP/LLNL was provided by the U.S.\ Department of Energy under contract
W-7405-ENG-48. DJH thanks the Institute of Astronomy, University of Cambridge
and the Raymond and Beverly Sackler Foundation for hospitality and support
during the completion of this work. This is contribution number 667 of the
Columbia Astrophysics Laboratory.

\sec {} {REFERENCES}

\bigskip
\hi Armus, L., Heckman, T.\ M., Weaver, K.\ A., \& Lehnert, M.\ D.\ 1995,
ApJ, 445, 666

\hi Awaki, H., Ueno, S., Koyama, K., Tsuru, T., \& Iwasawa, K.\ 1996, PASJ,
48, 409

\hi Benn, C.\ R., Rowan-Robinson, M., McMahon, R.\ G., Broadhurst, T.\ J., \&
Lawrence, A.\ 1993, MNRAS, 263, 98

\hi Boller, T., Meurs, E.\ J.\ A., Brinkmann, W., Fink, H., Zimmermann, U.,
\& Adorf, H.-M.\ 1992, A\&A, 261, 57

\hi Bookbinder, J., Cowie, L.\ L., Krolik, J.\ H., Ostriker, J.\ P., \& Rees,
M.\ 1980, ApJ, 237, 647

\hi Bradt, H.\ V.\ D., \& McClintock, J.\ E.\ 1983, ARA\&A, 21, 13

\hi Comastri, A., Setti, G., Zamorani, G., \& Hasinger, G.\ 1995, A\&A, 296, 1

\hi Condon, J.\ J.\ 1989, ApJ, 338, 13

\hi Dahlem, M., Hartner, G.\ D., \& Junkes, N.\ 1994, ApJ 432, 598

\hi Dahlem, M., Heckman, T.\ M., Fabbiano, G., Lehnert, M.\ D., \& Gilmore, D.\
1996, ApJ, 461, 724

\hi Dalton, W.\ W., \& Sarazin, C.\ L.\ 1995, ApJ, 440, 280

\hi David, L.\ P., Jones, C., \& Forman, W.\ 1992, ApJ, 388, 82

\hi Day, C., Arnaud, K., Ebisawa, K., Gotthelf, E., Ingham, J., Mukai, K.,
\& White, N.\ 1995, The ABC Guide to {\sl ASCA\/} Data Reduction
(Greenbelt:~NASA/GSFC)

\hi Della Ceca, R., Griffiths, R.\ E., \& Heckman, T.\ M.\ 1997, ApJ, 485, 581

\hi Della Ceca, R., Griffiths, R.\ E., \& Heckman, T.\ M., \& MacKenty,
J.\ W.\ 1996, ApJ, 469, 662

\hi Edmunds, M.\ G., \& Pagel, B.\ E.\ J.\ 1984, MNRAS, 211, 507

\hi Fabbiano, G.\ 1988, ApJ, 330, 672

\hi Fabbiano, G.\ 1989, ARA\&A, 27, 87

\hi Fabian, A.\ C., \& Barcons, X.\ 1992, ARA\&A, 30, 429

\hi Feast, M.\ W., \& Robertson, B.\ S.\ C.\ 1978, MNRAS, 185, 31

\hi Fomalont, E.\ B., Windhorst, R.\ A., Kristian, J.\ A., \& Kellermann,
K.\ I.\ 1991, AJ, 102,~1258

\hi Gendreau, K.\ C., et al.\ 1995, PASJ, 47, L5

\hi Graham, J.\ R., Wright, G.\ S., Meikle, W.\ P.\ S., \& Joseph, R.\ D.\
1984, Nature, 310, 213

\hi Griffiths, R.\ E., \& Padovani, P.\ 1990, ApJ, 360, 483

\hi Haarsma, D.\ B., \& Partridge, R.\ B.\ 1998, ApJ, 503, L5

\hi Hargrave, P.\ J.\ 1974, MNRAS, 168, 491

\hi Hasinger, G., Burg, R., Giacconi, R., Schmidt, M., Tr\"umper, J., \&
Zamorani, G.\ 1998, A\&A, 329, 482

\hi Heckman, T.\ M.\ 1980, A\&A, 87, 152

\hi Heckman, T.\ M.\ 1998, in Origins, eds.\ C.\ E.\ Woodward, J.\ M.\ Shull,
\& H.\ A.\ Thronson, ASP Conference Series, 148, 127

\hi Heckman, T.\ M., Armus, L., \& Miley, G.\ K.\ 1990, ApJS, 74, 833

\hi Heckman, T.\ M., Dahlem, M., Eales, S.\ A., Fabbiano, G., \& Weaver, K.\
1996, ApJ, 457,~616

\hi Heckman, T.\ M., Dahlem, M., Lehnert, M.\ D., Fabbiano, G., Gilmore, D.,
\& Waller, W.~H.\ 1995, ApJ, 448, 98

\hi Helfand, D.\ J., \& Moran, E.\ C.\ 1999, ApJ, in preparation

\hi Helou, G., Khan, I.\ R., Malek, L., \& Boehmer, L.\ 1988, ApJS, 68, 151

\hi Helou, G., Soifer, B.\ T., \& Rowan-Robinson, M.\ 1985, ApJ, 298, 7

\hi Joseph, R.\ D., \& Wright, G.\ S.\ 1985, MNRAS, 214, 87

\hi Kotilainen, J.\ K., Moorwood, A.\ F.\ M., Ward, M.\ J., \& Forbes,
D.\ A.\ 1996, A\&A, 305, 107

\hi Kron, R.\ G., Koo, D.\ C., \& Windhorst, R.\ A.\ 1985 A\&A, 146, 38

\hi Lehnert, M.\ D., \& Heckman, T.\ M.\ 1994, ApJ, 426, L27

\hi Lehnert, M.\ D., \& Heckman, T.\ M.\ 1995, ApJS, 97, 89

\hi Lehnert, M.\ D., \& Heckman, T.\ M.\ 1996, ApJ, 462, 651

\hi Leitherer, C., \& Heckman, T.\ M.\ 1995, ApJS, 96, 9

\hi Leitherer, C., Robert, C., \& Drissen, L.\ 1992, ApJ, 401, 596

\hi Lutz, D., et al.\ 1996, A\&A, 315, L137

\hi Madau, P., Ghisellini, G., \& Fabian, A.\ C.\ 1994, MNRAS, 270, L17

\hi Mather, J.\ C., et al.\ 1990, ApJ, 354, L37

\hi McCarthy, P.\ J., Heckman, T., \& van Breugel, W.\ 1987, AJ, 92, 264

\hi Mihos, J.\ C., \& Hernquist, L.\ 1996, ApJ, 464, 641

\hi Miley, G.\ 1980, ARA\&A, 18, 165

\hi Moorwood, A.\ F.\ M., \& Oliva, E.\ 1994, ApJ, 429, 602

\hi Moran, E.\ C., Halpern, J.\ P., \& Helfand, D.\ J.\ 1994, ApJ, 433, L65

\hi Moran, E.\ C., \& Lehnert, M.\ D.\ 1997, ApJ, 478, 172

\hi Nagase, F.\ 1989, PASJ, 41, 1

\hi Nandra, K., \& Pounds, K.\ A.\ 1994, MNRAS, 268, 405

\hi Norris, R.\ P., \& Forbes, D.\ A.\ 1995, ApJ, 446, 594

\hi Ohashi, T., \& Tsuru, T.\ 1992, in Frontiers of X-ray Astronomy
(Tokyo:~Universal Academy Press), 435

\hi Osterbrock, D.\ E.\ 1989, Astrophysics of Gaseous Nebulae and Active
Galactic Nuclei (Mill Valley: University Science Books)

\hi Portegies Zwart, S.\ F., \& Verbunt, F.\ 1996, A\&A 309, 179

\hi Ptak, A., Serlemitsos, P., Yaqoob, T., Mushotzky, R., \& Tsuru, T.\ 1997,
AJ, 113, 1286

\hi Rephaeli, Y., Gruber, D., Persic, M., \& MacDonald, D.\ 1991, ApJ, 380, L59

\hi Reynolds, R.\ J.\ 1990, in IAU Symp.\ 114, The Interstellar Disk-Halo
Connection in Galaxies, ed.\ H.\ Bloeman (Dordrecht: Kluwer), 67

\hi Richards, E.\ A., Kellermann, K.\ I., Fomalont, E.\ B., Windhorst, R.\ A.,
\& Partridge, R.~B.\ 1998, AJ, 116, 1039

\hi Richards, E.~A.\ 1998, BAAS, 30, 1326

\hi Rieke, G.\ H., Lebofsky, M.\ J., Thompson,~R.~I., Low,~F.~J., \&
Tokunaga,~A.~T.\ 1980, ApJ, 238, 24

\hi Rigopoulou, D., et al.\ 1996 A\&A, 315, L125

\hi Sanders, D.\ B., Soifer, B.\ T., Elias, J.\ H., Madore, B.\ F., Matthews,
K., Neugebauer, G., \& Scoville, N.\ Z.\ 1988, ApJ, 325, 74

\hi Sargent, A.\ I., Sanders, D.\ B., \& Phillips, T.\ G.\ 1989, ApJ, 346, L9

\hi Sargent, W.\ L.\ W., \& Filippenko, A.\ V.\ 1991, AJ, 102, 107

\hi Schaaf, R., Pietsch, W., Biermann, P.\ L., Kronberg, P.\ P., \& Schmutzler,
T.\ 1989, ApJ, 336, 722

\hi Schmidt, M., et al.\ 1998, A\&A, 329, 495

\hi Steidel, C.\ C., Adelberger, K.\ L., Giavalisco, M., Dickinson, M., \&
Pettini, M.\ 1999, ApJ, in press

\hi Stewart, G.\ C., Fabian, A.\ C., Terlevich, R.\ J., \& Hazard, C.\ 1982,
MNRAS, 200, 61P

\hi Strickland, D.\ K., Ponman, T.\ J., \& Stevens, I.\ R.\ 1997, A\&A,
320, 378

\hi Suchkov, A.\ A., Balsara, D., Heckman, T.\ M., \& Leitherer, C.\ 1994,
ApJ, 430, 511

\hi Suchkov, A.\ A., Berman, V.\ G., Heckman, T.\ M., \& Balsara, D.\ S.\
1996, ApJ, 463, 528

\hi Sutherland, R.\ S., \& Dopita, M.\ A.\ 1993, ApJS, 88, 253

\hi Tanaka, Y., Inoue, H., \& Holt, S.\ S.\ 1994, PASJ, 46, L37

\hi Tenorio-Tagle, G., Mu\~noz-Tu\~n\'onm C., P\'erez, E., \& Melnick, J.\
1997, ApJ, 490, L179

\hi Thuan, T.\ X., \& Condon, J.\ J.\ 1987, ApJ, 322, L9

\hi Tucker, W.\ H.\ 1975, Radiation Processes in Astrophysics (Cambridge:
MIT Press), 169

\hi Vacca, W.\ D.\ 1994, ApJ, 421, 140

\hi van Paradijs, J.\ 1995 in X-ray Binaries, eds.\ W.\ H.\ G.\ Lewin,
J.\ van Paradijs, \& E.\ P.\ J.\ van den Heuvel (Cambridge: Cambridge
University Press), 536

\hi Veilleux, S., \& Osterbrock, D.\ E.\ 1987, ApJS, 63, 295

\hi Voges, W., et al.\ 1996, IAU Circ.\ 6420

\hi Wang, J., Heckman, T., \& Lehnert, M.\ D.\ 1997, ApJ, 491, 114

\hi Weedman, D.\ W.\ 1987, in Star Formation in Galaxies, ed.\
C.\ J.\ Lonsdale (NASA CP-2466), 351

\hi White, N., Swank, J., \& Holt, S.\ 1983, ApJ, 270, 711

\hi Windhorst, R.\ A., et al.\ 1995, Nature, 375, 471

\hi Windhorst, R.\ A., Fomalont, E.\ B., Partridge, R.\ B., \& Lowenthal,
J.\ D.\ 1993, ApJ, 405, 498

\hi Windhorst, R.\ A., Miley, G.\ K., Owen, F.\ N., Kron, R.\ G., \& Koo,
D.\ C.\ 1985, ApJ, 289, 494

\hi Wright, E.\ L., et al.\ 1994, ApJ, 420, 450

\hi Yaqoob, T., Serlemitsos, P.\ J., Ptak, A., Mushotzky, R., Kunieda, H., \&
Terashima, Y.\ 1995, ApJ, 455, 508

%% TABLE 1
\break
\topglue 0.75truein
\hskip 0.2truein{\vbox{\baselineskip 16truept
\halign{\hfil#\hfil\tabskip 1.5em& \hfil#\hfil& \hfil#\hfil& \hfil#\hfil& \hfil#\hfil& \hfil#\hfil \tabskip 0pt \cr
\multispan{6}\hfil\apj TABLE 1\hfil\cr\noalign{\vskip 4pt}
\multispan{6}\hfil\apj Fits to the {\sl ASCA\/} Spectra of NGC~3256\hfil\cr
\noalign{\vskip 1em\hrule\vskip 2pt\hrule\vskip 1em}
 Components&          &  $kT$ (keV)&                  $N_{\rm H}$&               \cr
 In Model& Component$^{\rm a}$& or $\Gamma$& ($\times 10^{21}$ cm$^{-2}$)& $A^{\rm b}$& Model $\chi^2$ ($\nu$)\cr
\noalign{\vskip 1em\hrule\vskip 1em}
1& R-S&                        0.78 &                      8.4 & 18.7 & 206.8 (82)~~\cr
\cr
2& R-S&                        4.14 &                      6.5 & ~5.0 &  96.7 (80) \cr
 & R-S&                        0.73 &               $^{\rm c}$ & 10.4 &            \cr
\cr
2& R-S&                     11.80~~ &                      0.0 & ~3.6 &  83.7 (79) \cr
 & R-S&                        0.68 &                      7.9 & 12.8 &            \cr
\cr
2&  PL&                        2.40 &                      3.3 & ~3.3 &  92.4 (80) \cr
 & R-S&                        0.81 &               $^{\rm c}$ & ~3.6 &            \cr
\cr
2&  PL& \errorb {1.67} {0.72} {0.45} & ~~~\errora {0.0} {3.3} {0.0} & ~1.2 &  80.7 (79) \cr
 & R-S& \errorb {0.75} {0.10} {0.16} & ~~~\errora {7.2} {2.1} {3.4} & ~9.9 &            \cr
\cr
3&  PL& \errorb {1.68} {0.84} {1.22} & ~~~\errora {7.9} {1.9} {3.2} & ~1.2 &  69.8 (78) \cr
 & R-S& \errorb {0.80} {0.16} {0.15} &               $^{\rm c}$ & 13.0 &            \cr
 & R-S& \errorb {0.29} {0.40} {0.14} &            1.0$^{\rm d}$ & ~1.1 &            \cr
\cr
\noalign{\hrule}
\noalign{\vbox{\hsize=5.5truein\vskip 1em
$^{\rm a}$ PL = power law; R-S = Raymond-Smith plasma; TB = thermal
brems-strahlung.

$^{\rm b}$ Component normalization at 1 keV.  For PL components, $A$ has
units of $10^{-4}$ photons cm$^{-2}$ s$^{-1}$ keV$^{-1}$; for R-S components,
$A$ is equal to [$10^{-18} / 4 \pi D^2$] $\int {n_e}\! \! ^2 dV$, where
$D$ is the distance to the source in cm, $n_e$ is the electron density in
cm$^{-3}$, and $V$ is the volume of the emitting region in cm$^3$.

$^{\rm c}$ Absorption column density same as that applied to first component
in this model.

$^{\rm d}$ Absorption column density fixed at the Galactic value.

{\apj Notes.}---Indicated errors are 90\% confidence limits for four
interesting parameters ($\Delta \chi^2 = 7.78$).  Solar abundances are
assumed for all R-S components.
}}}}}
\eject

%% TABLE 2
\topglue 0.75truein
\hskip 0.35truein{\vbox{\baselineskip 16truept
\halign{\hfil#\hfil\tabskip 1.5em& \hfil#\hfil& \hfil#\hfil& \hfil#\hfil& \hfil#\hfil& \hfil#\hfil& \hfil#\hfil& \hfil#\hfil& \hfil#\hfil \tabskip 0pt \cr
\multispan{9}\hfil\apj TABLE 2\hfil\cr\noalign{\vskip 4pt}
\multispan{9}\hfil\apj Fluxes and Luminosities for the Three-Component Model\hfil\cr
\noalign{\vskip 1em\hrule\vskip 2pt\hrule\vskip 1em}
& \multispan{2}\hfil 0.1--2.4 keV\hfil&\multispan{2}\hfil 0.5--2.0 keV\hfil&\multispan{2}\hfil 2.0--10.0 keV\hfil&\multispan{2}\hfil 0.1--10.0 keV\hfil\cr
\noalign{\vskip -4pt}
& \multispan{2}\hrulefill&\multispan{2}\hrulefill&\multispan{2}\hrulefill&\multispan{2}\hrulefill\cr
\noalign{\vskip -2pt}
Component & \FX $^{\rm a}$ & \LX $^{\rm b}$  & \FX $^{\rm a}$ & \LX $^{\rm b}$ & \FX $^{\rm a}$ & \LX $^{\rm b}$ & \FX $^{\rm a}$ & \LX $^{\rm b}$\cr
\noalign{\vskip 1em\hrule\vskip 1em}
\omit PL \hfil         & 1.05 & 1.87 & 0.71 & 0.99 & 4.58 & 1.86 & 5.30 & 3.56 \cr
\cr
\omit R-S (warm) \hfil & 5.23 & 16.3~& 4.78 & 12.2~& 1.28 & 0.62 & 6.05 & 16.7~\cr
\cr
\omit R-S (soft) \hfil & 1.53 & 1.27 & 1.41 & 0.86 & 0.00 & 0.00 & 1.53 & 1.27 \cr
\cr
\omit Total \hfil      & 7.81 & 19.4~& 6.90 & 14.0~& 5.86 & 2.48 & 12.9~& 21.5~\cr
\cr
\noalign{\hrule}
\noalign{\vbox{\hsize=5.truein\vskip 1em
$^{\rm a}$ Observed X-ray flux in units of $10^{-13}$ ergs cm$^{-2}$ s$^{-1}$.

$^{\rm b}$ Intrinsic (unabsorbed) X-ray luminosity in units of $10^{41}$ ergs s$^{-1}$.
}}}}}
\eject

%% FIGURE CAPTIONS
\break
\sec {} {Figure Captions}

\bigskip
{\apj Fig.\ 1.}---Optical emission-line flux ratios as a function of offset
from the nucleus of NGC~3256 along PA = $65^{\circ}$, before ({\it filled
squares}) and after ({\it open squares}) correction for extinction and
Balmer absorption in the underlying stellar continuum.  Positive offsets
are west-southwest of the nucleus.

\bigskip
{\apj Fig.\ 2.}---Emission-line flux ratios as a function of offset from
the nucleus of NGC~3256 along PA = $155^{\circ}$, before ({\it filled
squares}) and after ({\it open squares}) correction for extinction and
Balmer absorption in the underlying stellar continuum.  Positive offsets
are north-northwest of the nucleus.

\bigskip
{\apj Fig.\ 3.}---Spectra of the nuclear ({\it upper panel}) and off-nuclear
({\it middle and lower panels}) regions of NGC~3256 along PA = $155^{\circ}$.
The line-flux ratios in the nuclear spectrum are typical of \HII\ regions.
The off-nuclear spectra, representing 8$''$ extractions centered 18$''$
(4.9 kpc) from the nucleus, exhibit enhancements of the low-ionization
\OI\ $\lambda 6300$, \NII\ $\lambda\lambda 6548,6583$, and
\SII\ $\lambda\lambda 6716,6731$ emission lines, similar to the ionized
gas observed in the halos of edge-on starburst galaxies.  The weak, broad
\Ha\ line present in the off-nuclear spectra is observed in {\it all\/}
extended emission-line regions sufficiently distant from the nucleus.

\bigskip
{\apj Fig.\ 4.}---Emission-line velocity widths as a function of offset
from the nucleus of NGC~3256 along PA = $155^{\circ}$.  The width of
\Ha\ does not include the contribution from a broad component (see Fig.~3),
if present.

\bigskip
{\apj Fig.\ 5.}---The \OI /\Ha\ flux ratio (corrected for both reddening and
Balmer absorption) vs.\ \Ha\ velocity width in NGC~3256. The apparent
correlation supports the notion that the off-nuclear emission in the galaxy
is the result of a shock-ionized wind.  The width of \Ha\ does not include
the contribution from a broad component (see Fig.~3), if present.

\bigskip
{\apj Fig.\ 6.}---Contours from the {\sl ROSAT\/} HRI image of NGC~3256,
overlayed on an optical image from the Digitized Sky Survey.  To emphasize
the low surface brightness X-ray emission, we binned the HRI image to have
$4''$ pixels and then smoothed it with a $\sigma_{\rm G} = 6''$ Gaussian.
The contours correspond to values of 1, 2.5, 4, 7, 11, and 15 (smoothed)
counts per pixel.  The axes are labeled with J2000 coordinates.  The soft
X-ray emission is confined to the main body of the galaxy, but extends
over a 12~kpc diameter.

\bigskip
{\apj Fig.\ 7.}---The 0.5--10 keV {\sl ASCA\/} spectrum of NGC~3256. ($a$)
The observed spectrum and best-fitting two-component (PL + R-S) model
(folded through the instrument response functions) are plotted in the
upper panel.  The unfolded spectrum shown in the lower panel indicates
that, in this model, the power-law component must dominate at both hard
and very soft X-ray energies.  In ($b$), the three-component model
from Table~1 has been applied.  The upper panel shows that this model
provides a slightly better fit than the two-component model.  As the
unfolded spectrum displayed in the lower panel indicates, the power-law,
warm (0.8~keV) thermal, and soft thermal components dominate in the hard,
medium-energy ($\sim$~0.8--2 keV), and very soft bands, respectively.

%% FIGURES
\break
{\topglue 1.0truein
\hsize 7.0truein
\hskip -0.5truein
\epsfxsize=6.5truein
\epsffile[27 162 563 690]{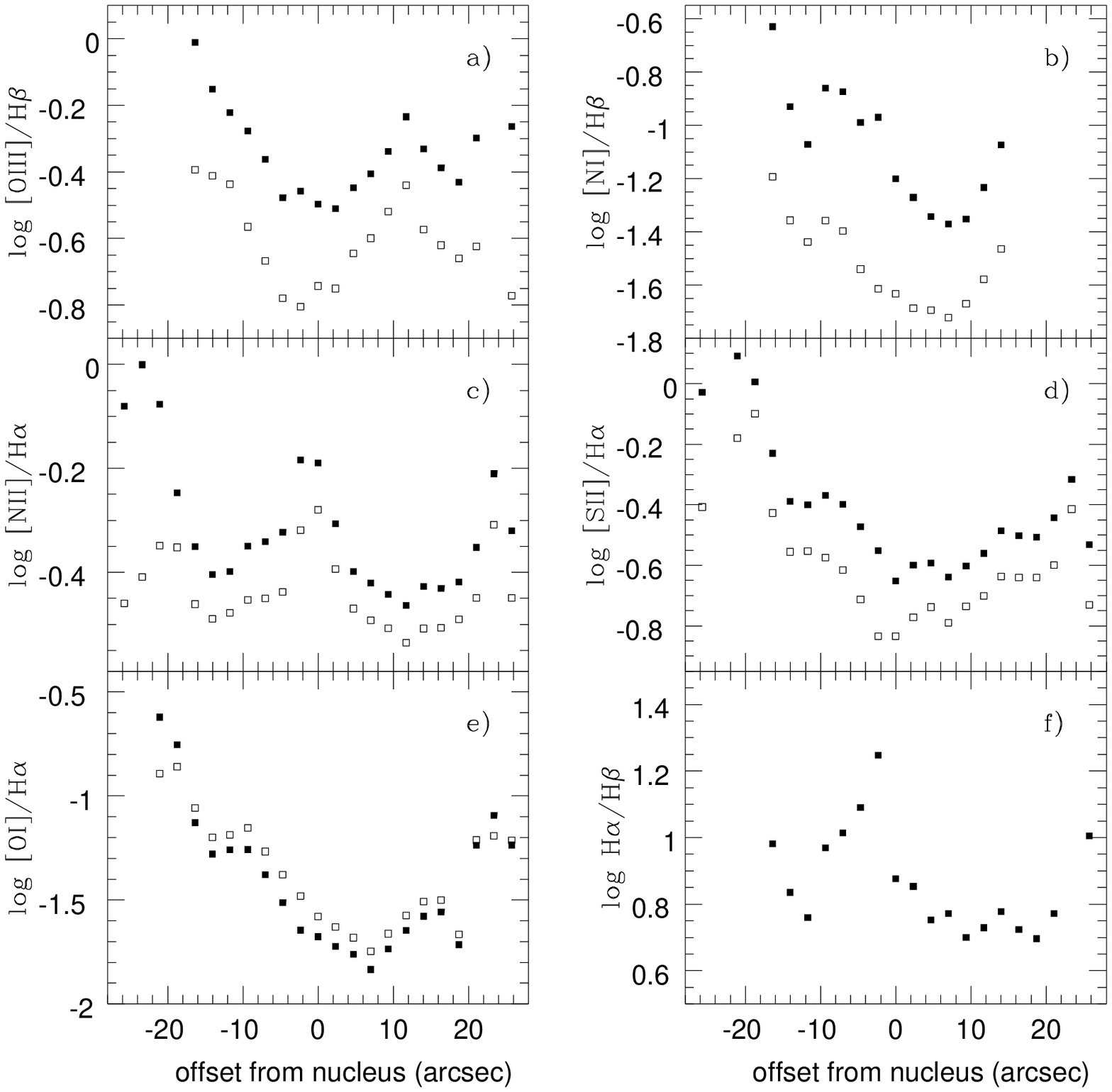}\par}

\vfill
\centerline{\apj Fig.~1}

\break
{\topglue 1.0truein
\hsize 7.0truein
\hskip -0.5truein
\epsfxsize=6.5truein
\epsffile[27 162 563 690]{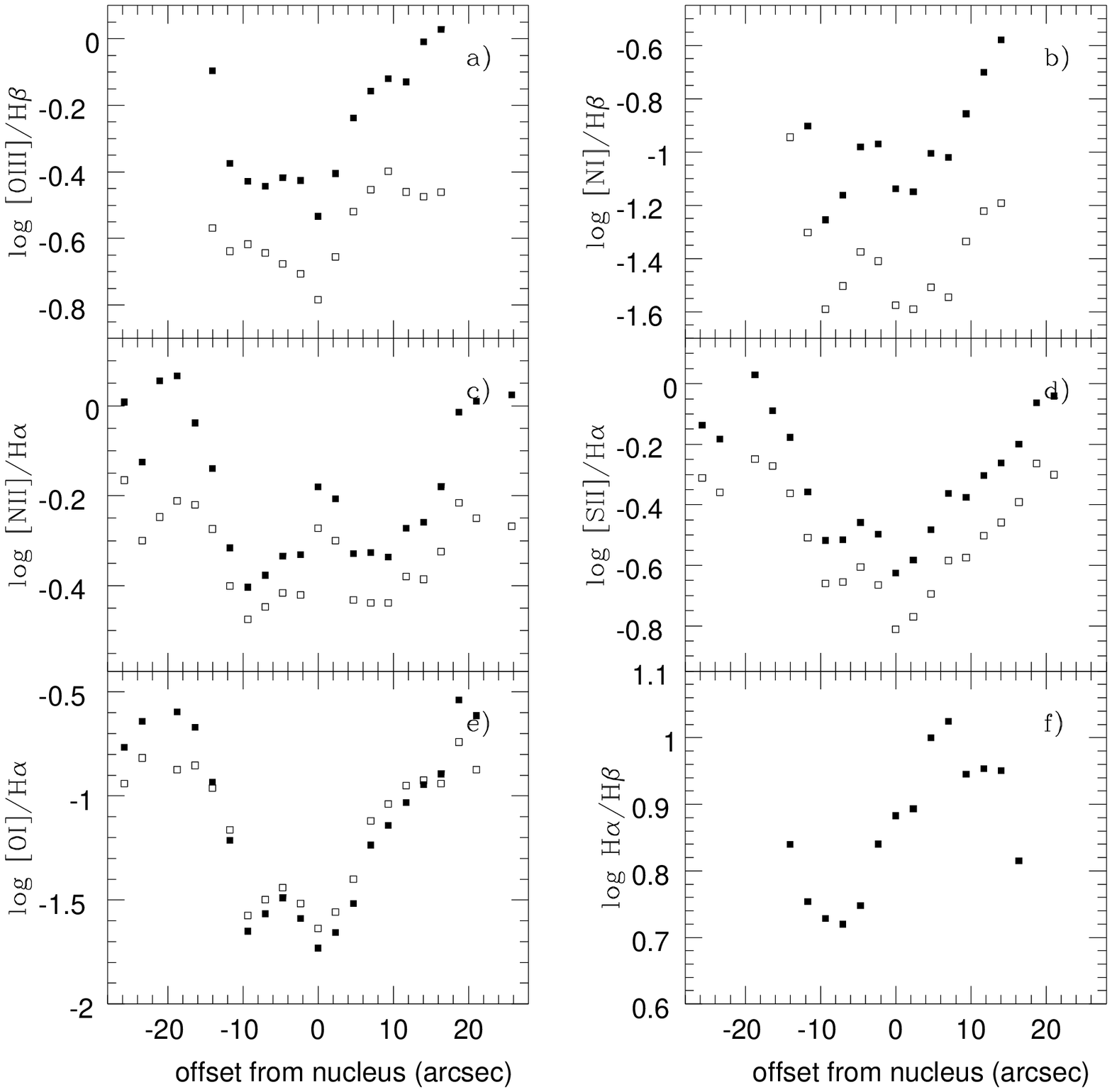}\par}

\vfill
\centerline{\apj Fig.~2}

\break
{\topglue 0.70truein
\hskip 0.2truein
\epsfysize=7.0truein
\epsffile[36 54 548 746]{fig3.ps}\par}

\vfill
\centerline{\apj Fig.~3}

\break
{\topglue 1.0truein
\hsize 7.0truein
\hskip -0.5truein
\epsfxsize=6.5truein
\epsffile[27 162 563 690]{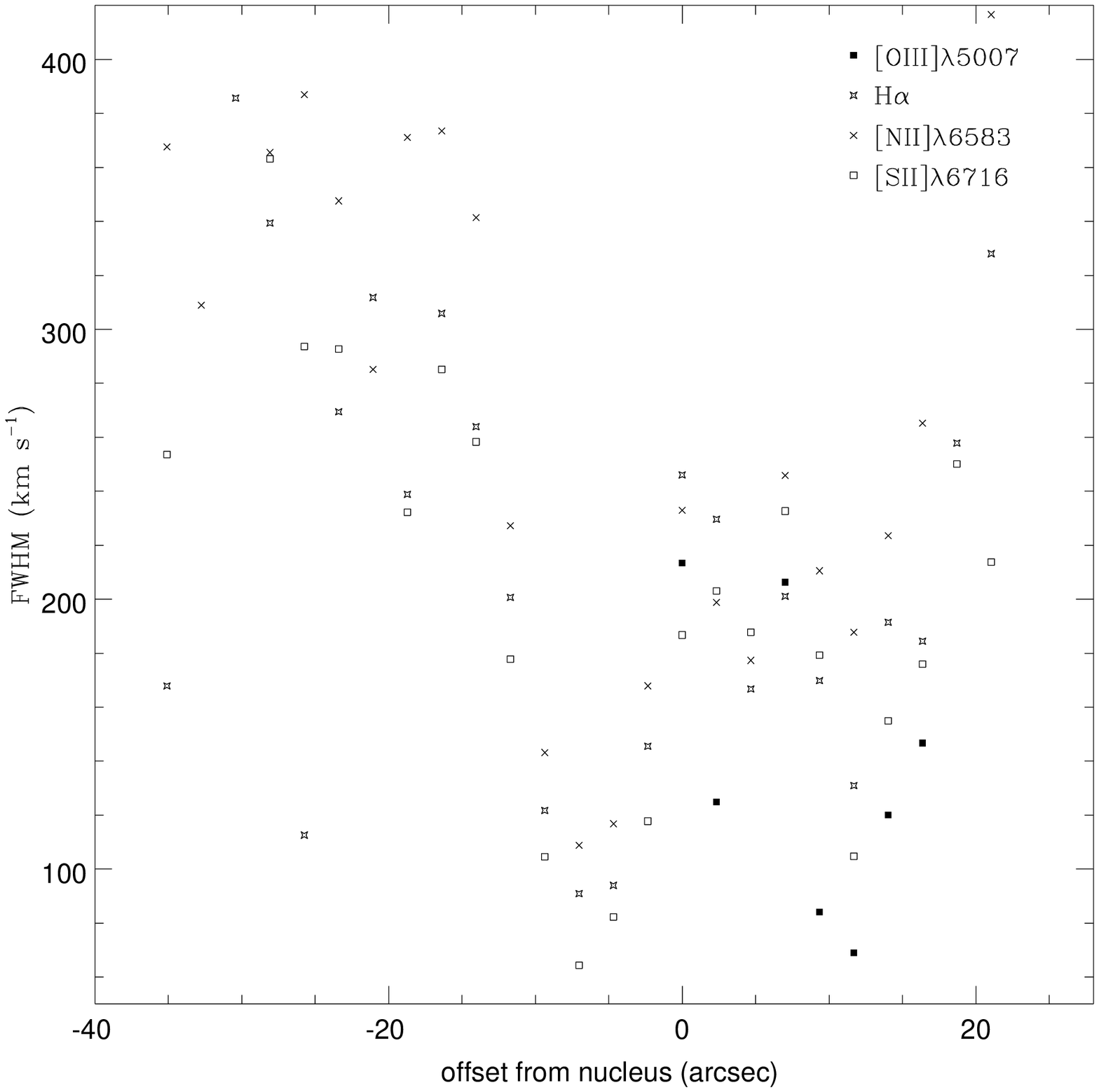}\par}

\vfill
\centerline{\apj Fig.~4}

\break
{\topglue 1.0truein
\hsize 7.0truein
\hskip -0.5truein
\epsfxsize=6.5truein
\epsffile[27 162 563 690]{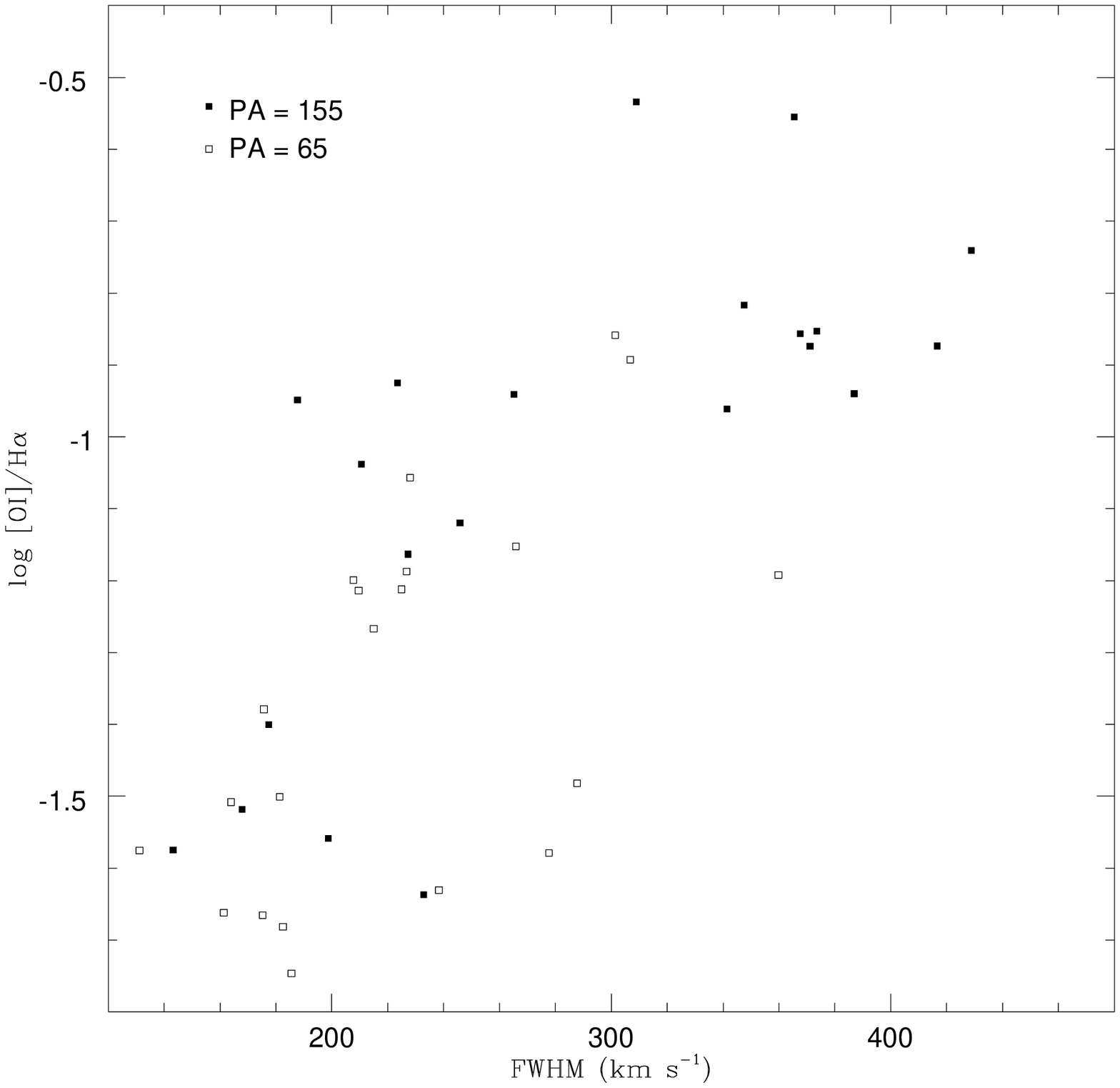}\par}

\vfill
\centerline{\apj Fig.~5}

\break
{\topglue 1.8truein
\hskip -0.4truein
\epsfxsize=6.0truein
\epsffile[26 220 580 630]{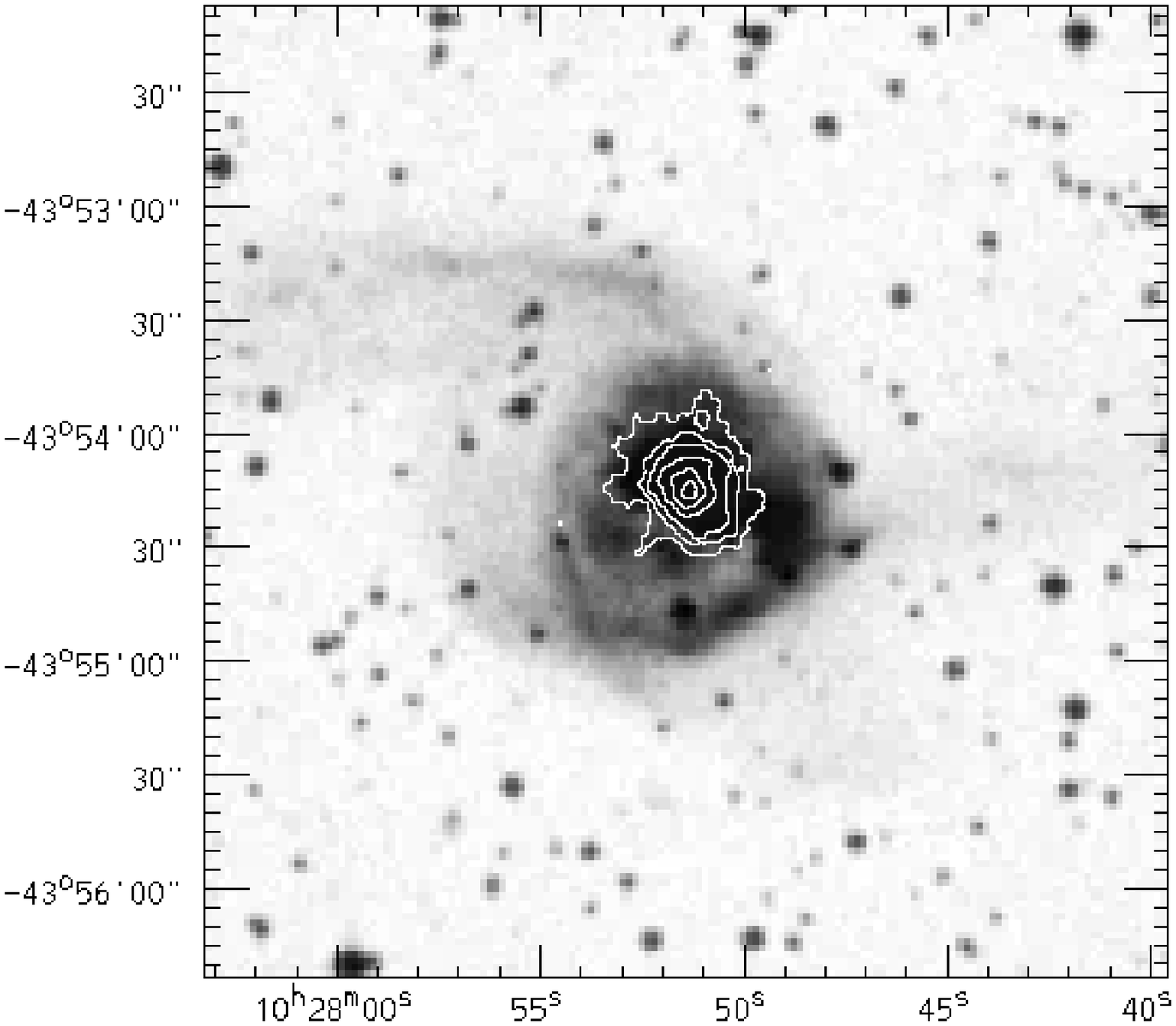}\par}

\vfill
\centerline{\apj Fig.~6}

\break
{
\hskip 0.6truein
\epsfysize=8.5truein
\epsffile[144 108 432 684]{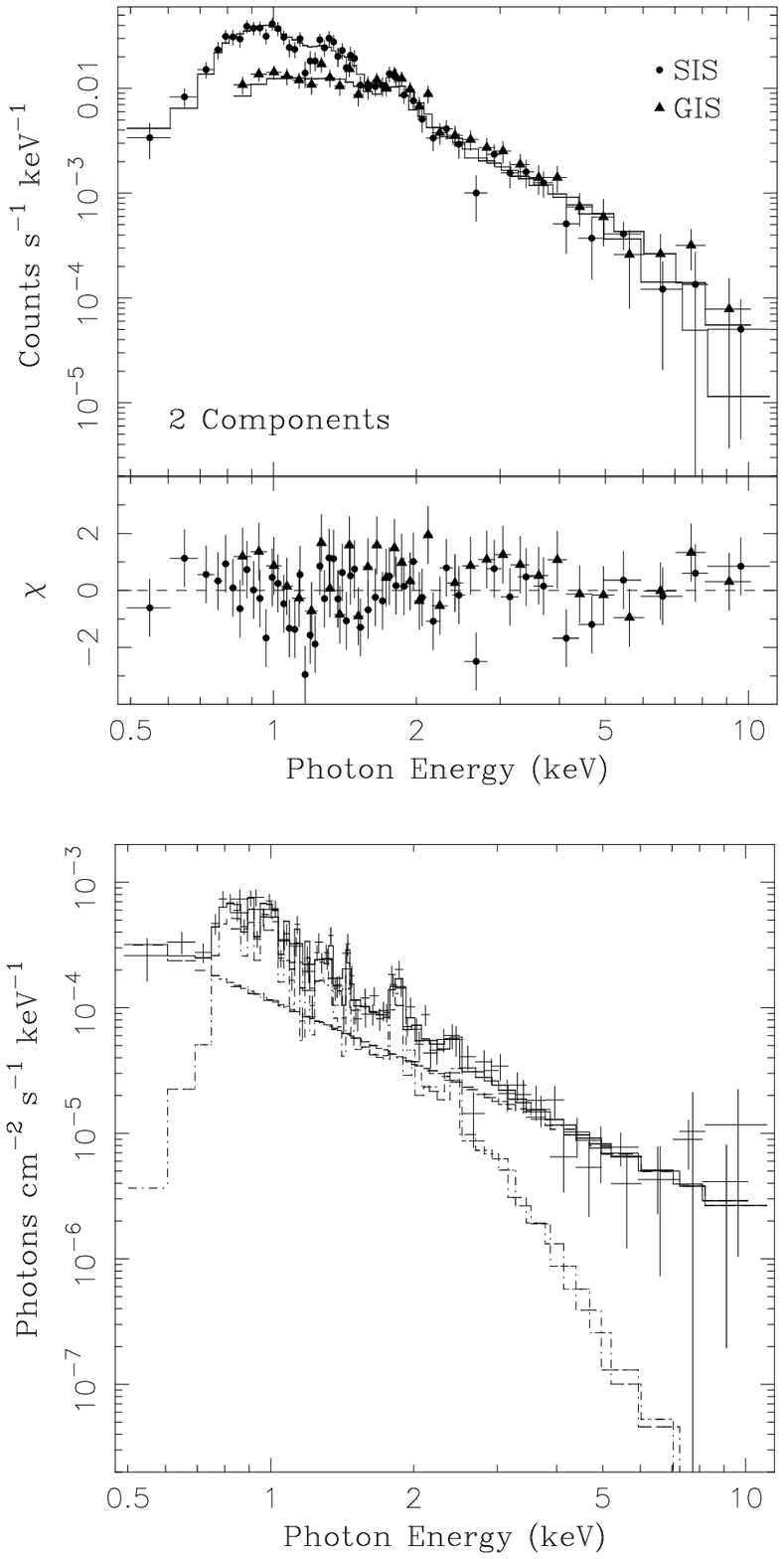}}

\vfill
\centerline{\apj Fig.~7$a$}

\break
{
\hskip 0.6truein
\epsfysize=8.5truein
\epsffile[144 108 432 684]{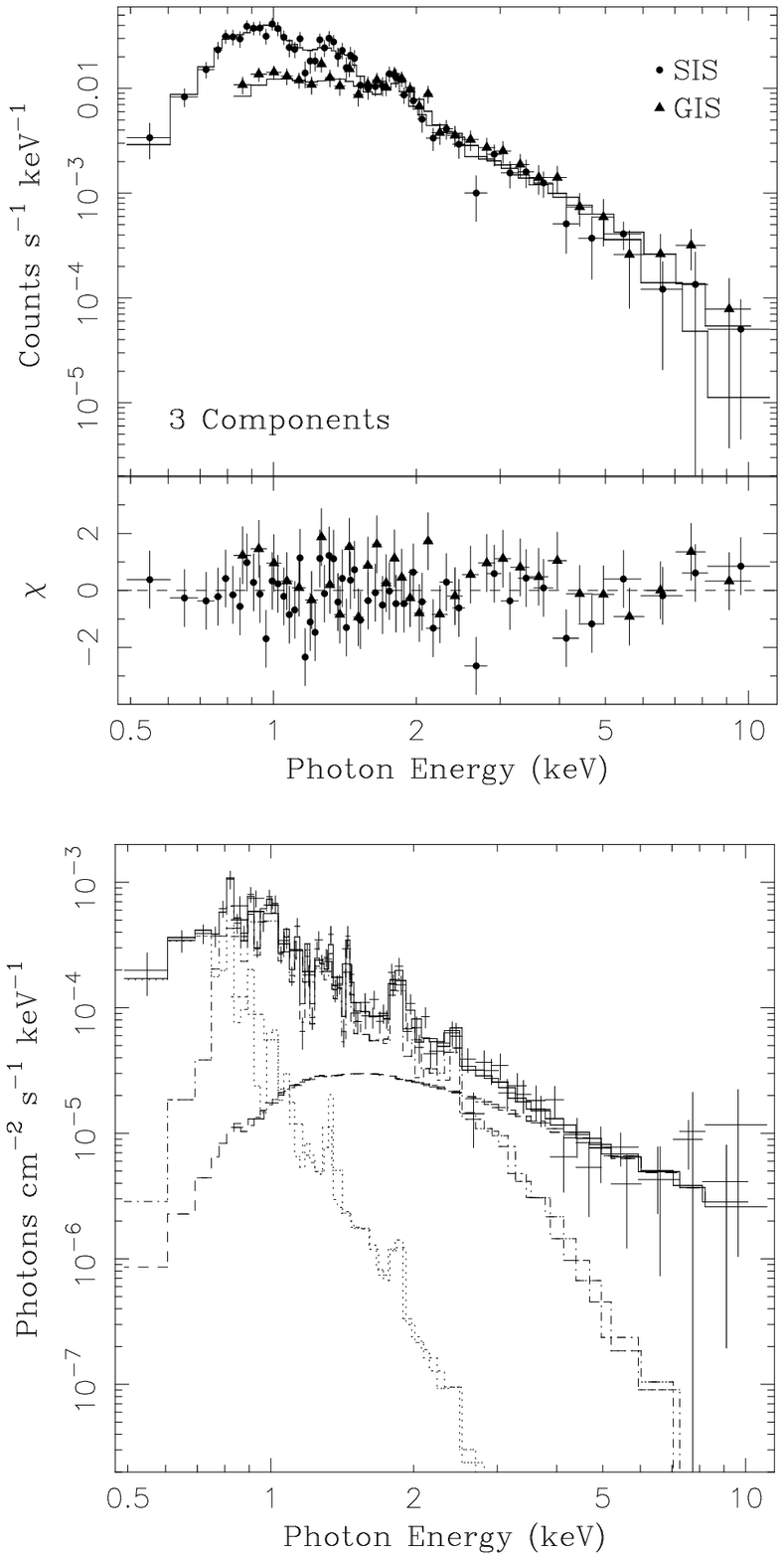}}

\vfill
\centerline{\apj Fig.~7$b$}

\bye